\documentclass[english,superscriptaddress,preprintnumbers,nofootinbib,amsmath,amssymb]{revtex4}
\usepackage{graphicx}
\usepackage{color}
\usepackage{hyperref}
\usepackage{bbm}
\newcommand{\be}{\begin{equation}}
\newcommand{\ee}{\end{equation}}
\newcommand{\bea}{\begin{eqnarray}}
\newcommand{\eea}{\end{eqnarray}}
\newcommand{\ben}{\begin{equation*}}
\newcommand{\een}{\end{equation*}}
\newcommand{\bean}{\begin{eqnarray*}}
\newcommand{\eean}{\end{eqnarray*}}
\newcommand{\bmat}{\begin{pmatrix}}
\newcommand{\emat}{\end{pmatrix}}
\newcommand{\bs}{\begin{split}}
\newcommand{\es}{\end{split}}

\newcommand{\f}{\frac}

\newcommand{\kaava}[1]{(\ref{#1})}

\newcommand{\im}{i}

\newcommand{\abs}[1]{\left\vert#1\right\vert}

\newcommand{\chq}{\textrm{ch}_q}

\newcommand{\Tr}{\textrm{Tr}}

\newcommand{\integers}{\mathbb{Z}}

\newcommand{\kd}[1]{\delta_{#1}}
\newenvironment{matriisi2}{\left[\begin{matrix}}{\end{matrix}\right]}
\newcommand{\qbinom}[2]{\begin{matriisi2} #1\\ #2\end{matriisi2}}

\newcommand{\bK}{\mathbf{K}}
\newcommand{\bl}{\mathbf{l}}

\newcommand{\bm}{\mathbf{m}}
\newcommand{\bn}{\mathbf{n}}

\begin{document}


\title{

Local height probabilities in a composite Andrews-Baxter-Forrester model}

\author{Jaakko Nissinen}
\affiliation{%
Theory group, Department of Physics, University of Oslo, NO-0316 Oslo, Norway
}
\affiliation{%
Nordita, Royal Institute of Technology and Stockholm University,
Roslagstullsbacken 23,
SE-106 91 Stockholm,
Sweden
}

\author{Eddy Ardonne}
\affiliation{%
Nordita, Royal Institute of Technology and Stockholm University,
Roslagstullsbacken 23,
SE-106 91 Stockholm,
Sweden
}
\affiliation{%
Department of Physics, Stockholm University,
AlbaNova University Center, SE-106 91 Stockholm, Sweden
}
\begin{abstract}
We study the local height probabilities in a composite height model, derived from the restricted solid-on-solid model introduced by Andrews, Baxter and Forrester, and their connection with conformal field theory characters. The obtained conformal field theories also describe the critical behavior of the model at two different critical points. In addition, at criticality, the model is equivalent to a one-dimensional chain of anyons, subject to competing two- and three-body interactions. The anyonic-chain interpretation provided the original motivation to introduce the composite height model, and by obtaining the critical behaviour of the composite height model, the critical behaviour of the anyonic chains is established as well. Depending on the overall sign of the hamiltonian, this critical behaviour is described by a diagonal coset-model, generalizing the minimal models for one sign, and by Fateev-Zamolodchikov parafermions for the other.
\end{abstract}
\date{\today}
\maketitle


\section{Introduction}

Ever since the advance of conformal field theory in the seminal paper by Belavin, Polyakov and Zamolodchikov \cite{bpz84}, it has played
an extremely important role in the study of critical behaviour in two-dimensional statistical mechanics models, and one-dimensional
quantum systems alike. Not only were the foundations of conformal field theory (CFT) laid out in \cite{bpz84}, in addition, an infinite
series of conformal field theories were introduced, the so-called minimal models. These CFTs describe an infinite number of possible
critical points. Roughly at the same time, Andrews, Baxter and Forrester \cite{abf84} studied a generalization of the eight-vertex model,
in which the degrees of freedom are heights, living on the square lattice. These heights can take a finite set of $r-1$ values, where $r$
is a parameter characterizing the model. Because of this constraint, these models are also called `restricted-solid-on-solid' (RSOS), or
simply `height' models. These models were shown to exhibit various gapped phases, separated by critical points. Shortly afterwards
Huse \cite{h84} realized that the critical points found by Andrews, Baxter and Forrester are described by the family of unitary minimal models obtained by Friedan, Qiu and Shenker around the same time \cite{fqs84}. One can certainly say that conformal field theory took a flying start! Here we study aspects of a similar connection between critical points in composite height models, as well as in anyonic quantum chains, and CFTs. 

Specifically, we study the local height probabilities (LHPs), to be defined below, of generalized RSOS models \cite{ka12} inspired by an anyonic quantum chain with competing two- and three-body interactions, which was introduced in \cite{tafhlt08}. Originally \cite{f07}, the anyonic chains were motivated as simple models for interacting anyons in topological phases, hosting the anyons constituting the chain as their elementary excitations. Indeed, the local degrees of freedom of the anyon chain related to the height model we study here are non-abelian $su(2)_{k}$ anyons and --- exactly as for a Heisenberg chain of $su(2)$ spins --- the anyon hamiltonian assigns an energy cost depending on to which representation the two or three neighboring anyons are `fused'. In contrast to spin chains, however, only the $k+1$ integrable representations of $su(2)_{k}$ can appear and, crucially, the Hilbert space of the chain is not a local tensor product of the single anyon degrees of freedom. Previous studies have revealed that the anyonic chains have very rich phase diagrams, even richer than in the original spin models, with novel phases protected or broken by `topological symmetries' \cite{f07, tafhlt08}. Here we study two integrable critical points of the anyon model with competing two- and three-body interactions, identified in \cite{ka12}. These integrable points of the anyon chain are directly obtained from an `anyonic' representation of the Temperley-Lieb algebra and are equivalent with the critical points of a classical, integrable `composite' RSOS height model also introduced in \cite{ka12}. This mapping is crucial, since the non-local form of the Hibert space of the anyon chain reduces the utilizability of Bethe ansatz techniques (but see \cite{br90,w96a}).

The RSOS height model is defined on a square lattice with the heights $l_{i}$ being local degrees of freedom on each vertex $i$, subjected to the constraints $1\leq l_{i} \leq r-1$ with $\abs{l_{i}-l_{j}}=1$ for adjacent vertices $i,\ j$, and $r=k+2$ for $su(2)_{k}$ anyons, and is a composite of the original integrable RSOS models of Andrews, Baxter and Forrester (ABF) \cite{abf84}. The various critical points of the ABF model, separating the ordered phases, were subsequently identified in \cite{h84}, and shown to provide realizations of the minimal models studied in the seminal papers \cite{bpz84,fqs84}. This type of multi-critical behavior also applies to the composite model. The study of the critical, as well as off-critical behavior of the model goes via the so-called local height probabilities, which are the probabilities for a central site to have a certain height, given the boundary conditions. As with the original ABF model, the off-critical LHPs of the height model exhibit properties related to conformal field theory characters and can be calculated exactly with the corner transfer matrix method \cite{saleur88, date86, date87}. This off-critical CFT structure of the LHPs is governed by the same theory that describes the critical points of the lattice model; the former arises from integrable perturbations of the latter. In particular, for a finite lattice size, the LHPs are composed of finitized forms of CFT characters from which one can obtain the characters by taking the thermodynamic limit.

Thus the two integrable points of the anyon chain are related to two different regimes of the height model, regimes II and III in the notation of \cite{abf84, ka12}, not just the critical points to which they terminate. Therefore LHPs allow one to determine the (extended) critical behavior of the anyon chain once the off-critical CFT has been identified, which is the aim of the current paper.
 
The central objects of interest in our paper are fermionic generating functions, quantities usually called universal chiral partition functions (UCPFs) \cite{bmc98}, or fundamental fermionic forms \cite{bmcs98, welsh05}, which we want to reproduce the LHPs in closed form and relate to finitized forms of CFT characters.

As with the first proofs of the connection with the ABF model and the $\mathcal{M}(r-1,r)$ minimal models, our strategy of proof is based on the recurrence properties of the polynomial UCPFs and LHPs, this strategy is sometimes referred to in the literature as Schur's method. In the thermodynamic limit, our formulas give fermionic characters of the coset conformal field theory $\f{su(2)_{1}\times su(2)_{1}\times su(2)_{r-4}}{su(2)_{r-2}}$ in the regime III and $Z_{r-2}$ parafermions in the regime II, corresponding to the two integrable points under study. Based on numerical checks \cite{tafhlt08, ka12}, these CFTs were earlier identified as the critical behavior of the anyonic chain. The parafermion theory and characters are well known and appear also in the ABF models; we recover the fermionic forms of these characters \cite{lp85, kkmcm93, kkmcm93a, g95up, jacob02}. The fermionic characters we obtain for the coset theory are new to the best of our knowledge.

Finally, the fermionic forms of the LHPs of the composite model studied here open an arena of $q$-identities related to the coset theories $\f{su(2)_{1}\times su(2)_{1}\times su(2)_{r-4}}{su(2)_{r-2}}$. This is exactly like the $q$-identity and CFT character results obtained for the minimal models from the ABF-type models \cite{abf84, fb85, kkmcm93, kkmcm93a, welsh05, melzer94,melzer94b, berkovich94, s96, w96b, bmcs98, foda99} and provides another motivation to study the composite height model. The UCPFs that we obtain in the regime III for the coset theory are characterized by the fact that they have two `real' fermions and $r-5$ `pseudo' particles, instead of just one `real' fermion appearing in the ABF type expressions \cite{melzer94, berkovich94, s96, w96b}. Given that one would be able to obtain the bosonic forms of the characters, giving interesting bose-fermi type identies, one could possibly also obtain new types of Rogers-Ramanujan and $q$-identities \cite{abf84, fb85, w96a, w96b, bmcs98, welsh05}. As already mentioned, the dual finitized characters obtained for the regime II are $Z_{r-2}$ parafermions, as in the ABF case, with all fermions `real'. In fact, this type of behavior seems to be rather generic, as our calculations in sec. \ref{sec:parafermions} suggest.

This paper is organized as follows. In section \ref{sec:heightmodel} we briefly introduce and recollect the composite height model from Ref. \cite{ka12} and set out the stage and notation for the various quantities related to the LHPs. As in the height model of ABF, the composite model is parametrized by an integer $r$ that sets the maximum of the height variables, with  $r\geq5$ for the anyonic chains.  In the paper \cite{ka12}, a fermionic form for the central quantity $X_{m}$ in the LHPs for the simplest case $r=5$ was introduced, where $m$ is the lattice size of the height model, based on numerical checks. In section \ref{sec:r=5}, we prove this equality analytically and then proceed for the $r=6$ case in sec. \ref{sec:r=6}. In both cases, the proof is obtained using the recurrence properties of the LHPs and the fermionic generating functions related to those with respect to the lattice size, exactly as in the original ABF model. In section \ref{sec:conjecture}, based on the structure for $r=5,6$, we give the general form of our fermionic UPCFs related to the LHPs, a conjecture we claim valid for any $r$ and supported by numerical checks for $r > 6$ and correct central charges. While we could apply the same the strategy of proof for bigger $r$, the computations become quickly cumbersome and not very illuminating. In sections \ref{sec:thermodynamic} and \ref{sec:CFT}, we finally study the thermodynamic limit of the UPCFs and give the explicit connection with CFT characters, respectively. We end by discussing our results and giving some future directions of study.

\section{Local height probabilities in a composite height model}{\label{sec:heightmodel}}

The height model we study in this paper is most easily explained in terms of the
original height model introduced by Andrews, Baxter and Forrester (ABF) \cite{abf84}.
We will first introduce this model, and subsequently explain how the composite
height model can be constructed from it.

\subsection{Definition of the height models}

\begin{figure}[hb]
\includegraphics{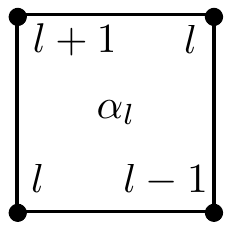}
\hspace{0.5cm}
\includegraphics{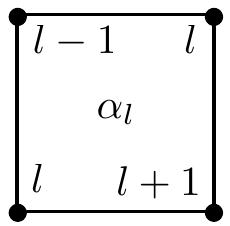}
\hspace{0.5cm}
\includegraphics{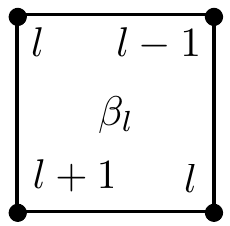}
\hspace{0.5cm}
\includegraphics{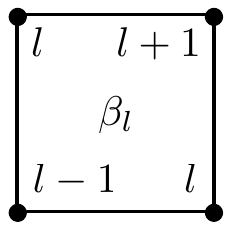}
\hspace{0.5cm}
\includegraphics{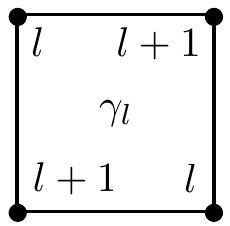}
\hspace{0.5cm}
\includegraphics{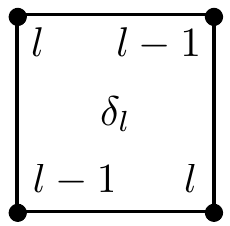}
\caption{The six different type of plaquettes occurring in the ABF model.}
\label{fig:abfplaquettes}
\end{figure}

The ABF model consists of heights which are assigned to the vertices of the square
lattice. The heights $l$ can take the values $l=1,2,\ldots,r-1$, where $r$ is an integer
satisfying $r\geq 3$. Different values of $r$ correspond to different models; the case
$r=5$ is equivalent to the hard hexagon model, as explained in \cite{abf84}.
Heights at neighbouring vertices have to satisfy the constraint
that they differ by one. This constraint leads to six different types of plaquettes on the
square lattice, as depicted in Fig.~\ref{fig:abfplaquettes}.
The two plaquettes labeled by $\alpha_l$ (and similarly for $\beta_l$) will be assigned the
same weight, so that one obtains an isotropic model.
ABF showed that this model can be solved for a two-parameter family of weights. We will
denote these parameters by $p$ and $u$. The first parameter $-1\leq p \leq 1$
resembles a temperature, and drives a phase transition at $p=0$. The parameter $u$
is related to the `anisotropy' of the lattice, and is the variable appearing in the Yang-Baxter
equation below. The behaviour of the model will not depend on
the magnitude of $u$, only its sign. For a description of the various phases of the ABF model,
we refer to the original paper \cite{abf84}, and the paper by Huse \cite{h84}, who studied the
connection between the critical points of the model and conformal field theory. We will come
back to the various phases of the composite height model after we explained how the model
can be solved.

The weights for which the ABF model can be solved explicitly are given in terms of
elliptic functions. To specify them, we introduce the following notation. First, $p$ will be
related to the modulus $m^2$ of the theta functions via
$p = e^{-\pi\frac{K'(m)}{K(m)}}$, where $K(m)$ is the complete elliptic integral of the first
kind and $K'(m) = K(1-m)$.
The parameter $r$ enters the weights via $\eta = \frac{K(m)}{r}$, while the values
of the heights $l$ enter as $w_l = 2 \eta l$.

Introducing the elliptic functions $H(u) = \theta_{1} (\frac{u\pi}{2K(m)},p)$
and $\Theta(u) = \theta_{4} (\frac{u\pi}{2K(m)},p)$, we define
$h(u) = H(u) \Theta(u)$, where we have suppressed the dependence on $p$ (or $m$), as
is customary. The $\theta_{i}$ are the Jacobi theta functions.
Explicitly, one finds the following expression for $h(u)$,
\begin{equation}
h(u) = 2 p^{\frac{1}{4}} \sin\bigl( \frac{\pi u}{2K} \bigr)
\prod_{n=1}^{\infty}
(1-2p^n \cos\bigl( \frac{\pi u}{K} \bigr) +p^{2n})
(1-p^{2n})^2 \ .
\end{equation}  
We can now introduce the two-parameter family of plaquette weights as follows
\begin{align}
\label{eq:abfweights}
\alpha_l (u) &= \frac{h(2\eta -u)}{h(2\eta)} &
\beta_l (u) &= \frac{h(u)}{h(2\eta)} \frac{\bigl( h(w_{l-1}) h(w_{l+1}) \bigr)^{\frac{1}{2}}}{h(w_l)} \\
\nonumber
\gamma_l (u) & = \frac{h(w_{l}+u)}{h(w_l)} &
\delta_l (u) & = \frac{h(w_{l}-u)}{h(w_l)} \ .
\end{align}

As a first step in solving their model, ABF noted that the weights \eqref{eq:abfweights} satisfy
the Yang-Baxter equation, which implies that the row-to-row transfer matrices for different
values of $u$ commute with each other.
We will explain in a bit more detail how the model
was solved below, after we introduced the composite height model considered in
\cite{ka12}.

\begin{figure}[t]
\includegraphics{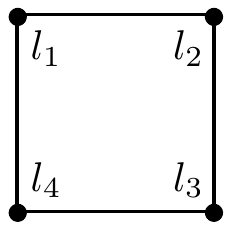}
\caption{The plaquettes of the ABF model.}
\label{fig:plaquette}
\end{figure}

Inspired by the the work of Ikhlef {\em et al.} \cite{ijs09,ijs10} on loop models, a composite model was
constructed in the following way. First, we denote the weights associated with a
plaquette by $W(u;l_1,l_2,l_3,l_4)$ as depicted in Fig.~\ref{fig:plaquette}.

\begin{figure}[b]
\includegraphics{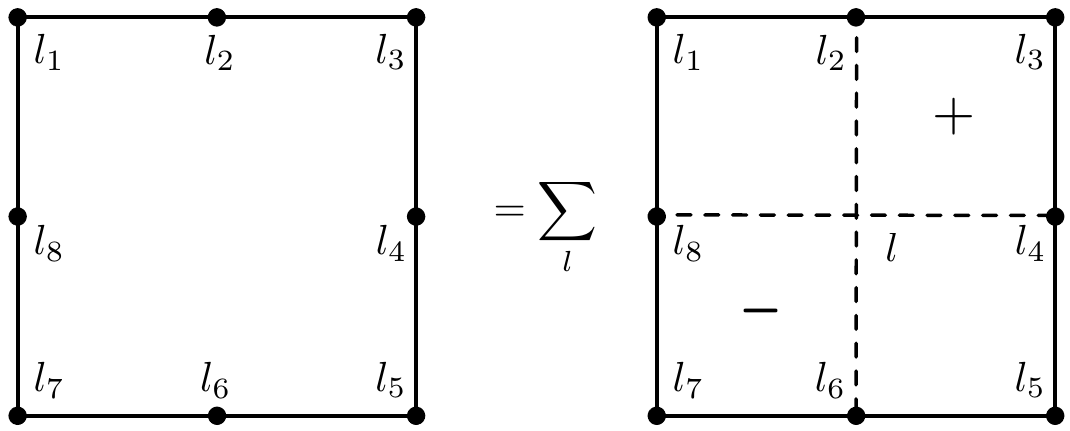}
\caption{The plaquette weights of the composite model, where the symbols $\pm$ denote the corresponding shifts of $u$ in \eqref{eq:compositeW}.}
\label{fig:comp-plaquette}
\end{figure}

By taking four of these $(2\times 2)$ plaquettes, one can from a composite $(3\times 3)$
plaquette, with the weight (see Fig.~\ref{fig:comp-plaquette})
\begin{equation}
\widetilde{W} (u;l_1,l_2,l_3,l_4,l_5,l_6,l_7,l_8) =
\sum_{l}
W(u;l_1,l_2,l,l_8) 
W(u+K;l_2,l_3,l_4,l) 
W(u;l,l_4,l_5,l_6) 
W(u-K;l_8,l,l_6,l_7) \ . \label{eq:compositeW}
\end{equation}
We note that the parameter $u$ of two of the sub-plaquettes has been shifted. Without
this shift, the composite model would be equivalent to the original model. It is a straightforward
exercise to show that the plaquette weights $\widetilde{W}$ of the composite
model satisfy the Yang-Baxter equation, by only making use of the fact that the plaquette weights
$W$ of the original model satisfy the Yang-Baxter equation, namely
\begin{equation}
\sum_{l} W(u;l_1,l,l_5,l_6) W(u+v;l_2,l_3,l,l_1) W(v;l_3,l_4,l_5,l) =
\sum_{l} W(v;l_2,l,l_6,l_1) W(u+v;l,l_4,l_5,l_6) W(u;l_2,l_3,l_4,l) \ .
\label{eq:yb}
\end{equation}

\subsection{A glimpse on the Corner Transfer Matrix method}

We now briefly discuss how these height models can be solved.
One makes use of the Corner Transfer Matrix (CTM) method, which was explained
in detail in chapter 13 of Baxter's book \cite{book:b82} (see also \cite{b07} for a recent
account). The application of the CTM method to solve the ABF model is detailed in
\cite{abf84}, while \cite{ka12} deals with the composite model.

The key objects in the CTM method are four corner transfer matrices. In contrast to the
row-to-row transfer matrix, the corner transfer matrices do not add merely one
row to the lattice, but instead an entire quadrant, or corner. If we denote the CTMs of the
four different corners by $A$, $B$, $C$ and $D$, the partition function of the
model is given by $Z = \Tr (ABCD)$. Using the CTM method, one can calculate a
quantity called the local height probability (LHP). In particular, $P_a$ denotes the
probability for the central height, say $l_1$ (at the corner of the four CTMs)
to take the value $a$. This probability can be written as
\begin{equation}
P_a = \frac{1}{Z} \Tr (S_a A B C D) \ ,
\end{equation}
where $S_a$ is the diagonal matrix, with diagonal entries $1$ if the central
height $l_1 =a$, and zero otherwise.

\begin{figure}[ht]
\includegraphics[height=5cm]{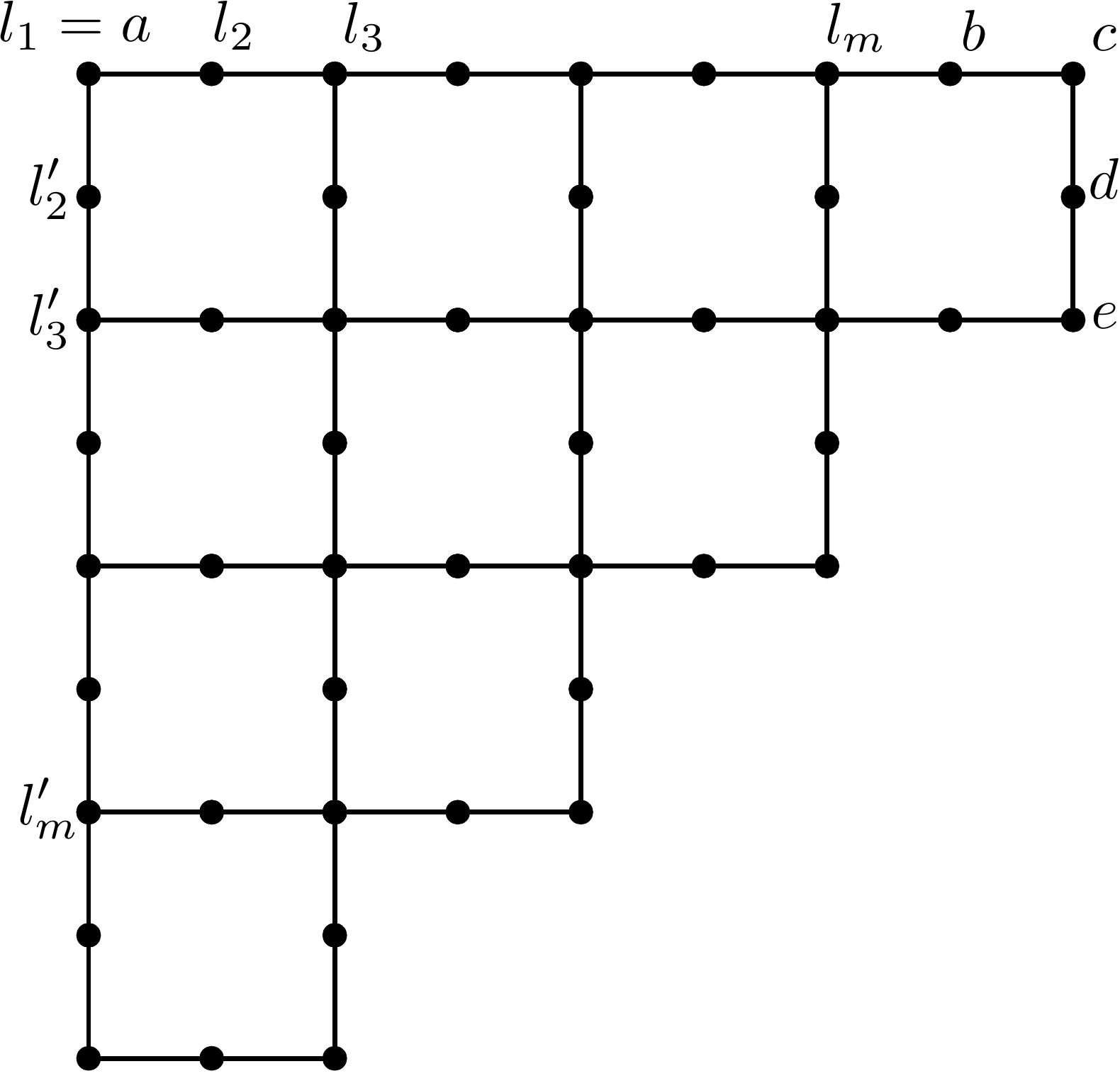}
\caption{The CTM $A$ of the composite model, with the boundary conditions $(b,c,d,e)$ as specified in the text.}
\label{fig:ctma}
\end{figure}

To make this discussion a bit more explicit, we display the CTM $A$ explicitly in
Fig.~\ref{fig:ctma}. The rows and columns of $A$ are labeled by
$(l_1, l_2, \ldots l_m)$ and $(l'_1 = l_1, l'_2, \ldots, l'_m)$ respectively.
The central height $l_1$ is fixed to be $a$, while the heights $l_{m+1}, l'_{m+1}, l_{m+2}, l'_{m+2}$, etc., at the boundary are fixed to ground state values of the model. By analyzing the weights of the model, one can show \cite{ka12} that the ground states are in fact diagonal, in this case along the SW-NE direction, and are fixed by the boundary conditions $(b,c,d,e)$ as indicated in the figure. The different ground state `patterns' $(b,c,d,e)$ are discussed in detail in section \ref{subsec:phases}, following \cite{ka12}. The CTM method allows one to calculate the
local height probabilities $P_a$, and these depend on the boundary conditions $(b,c,d,e)$.
The matrices $B$, $C$ and $D$ are obtained in a similar way as $A$, by subsequent
rotations of the diagram over $\pi/2$.
From the definition of $A$, it is clear that the 'size' of the quadrant $m$, equal to twice the number of the top-row plaquettes plus one, has to be odd.

We do not explain the calculation of the LHPs in full detail, but merely state the main
ingredients (following \cite{abf84, ka12}) of this calculation and of course the result, which
is an expression for the
height probabilities. These local height probabilities are the
starting point for the current paper, and the goal is to prove that the LHPs are
equal to (finitized) characters in conformal field theory (CFT).

The first essential ingredient of the CTM method is to make use of the Yang-Baxter
equation, to show that the CTMs $A$ etc. can be written in a special, diagonal form,
see \cite{book:b82} for the details.
One starts by formally equating the product $\lim_{m\rightarrow \infty}B(u)C(v)$,
which covers half of the lattice, to the limit $\lim_{n\rightarrow \infty} T(u,v)^n$ using the inhomogenous row-to-row transfer matrix $T(u,v)$, the latter limit also then covering half of the lattice with the anisotropies $u, v$ in the two quadrants \cite{book:b82}.
The Yang-Baxter equation can be shown to ensure that $B(u) C(v)$ only depends
on the difference $u-v$. In the end, one obtains the following form
for the CTMs
\begin{align}
A(u) &= Q_1 M_1 e^{-u \mathcal{H}} Q_2^{-1} \\ \nonumber
B(u) &= Q_2 M_2 e^{u \mathcal{H}} Q_3^{-1} \\ \nonumber
C(u) &= Q_3 M_3 e^{-u \mathcal{H}} Q_4^{-1} \\ \nonumber
D(u) &= Q_4 M_4 e^{u \mathcal{H}} Q_1^{-1} \ ,
\end{align}
where $\mathcal{H}$, $Q_i$ and $M_i$ (with $i=1,2,3,4$) do not depend on $u$, commute with the matrices $S_{a}$, and in addition $\mathcal{H}$ and $M_i$ are diagonal.

To relate these diagonal forms of the corner transfer matrices to the height probabilities,
one has to calculate the
form of the CTMs for particular values of $u$. In the case of the composite model,
one has to use an identity relating particular sums of products of elliptic
functions to a single product. The details can be found in \cite{ka12} but we summarize the results here. By the periodicity properties of the elliptic weights, one only needs to consider $u$ in the two following regimes
\be
\mathcal{D}_{1}: 0 < u < 2\eta + K = (2+r)\eta, \quad \mathcal{D}_{2}:2\eta-K = (2-r)\eta < u <0 .
\ee
One can show that, up to scalar multiples,
\be
A(0) = Q_{1}M_{1}Q_{2}^{-1} = \mathbbm{1},
\ee
which allows one to write
\be
A(u) = Q_{2}  e^{-u\mathcal{H}} Q_{2}^{-1},
\ee
so the diagonal form of $A(u)$ is equal to an exponential. The height probability $P_{a}$ is then given by
\be
P_{a}(b,c,d,e) = \f{\Tr(S_{a}M_{1}M_{2}M_{3}M_{4})}{\Tr(M_{1}M_{2}M_{3}M_{4})}.
\ee
and will depend in addition on the boundary conditions $(b,c,d,e)$. The product $M_{1}M_{2}M_{3}M_{4}$ can be computed by considering different limits of the corner transfer matrices.  First, in the limit $u\to 0$ in the domain $\mathcal{D}_{1}$ and up to irrelevant scalar factors, one has
\be
A(0) = C(0) = \mathbbm{1},
\ee
and secondly, when $u\to (2+r)\eta$, one has
\be
B(u= (2+r)\eta) = D(u= (2+r)\eta) = \widetilde{V}_{1},
\ee
where
\be
(\widetilde{V}_{1})_{\bl, \bl'} = \sqrt{h(2\eta l_{1})}\delta(\bl,\bl').
\ee
Therefore,
\be
A(0)B((2+r)\eta)C(0)D((2+r)\eta) = M_{1}M_{2}M_{3}M_{4}e^{2(2+r)\eta} = \widetilde{V}_{1}^{2}.
\ee
Similarly in the domain $\mathcal{D}_{2}$, where the weights effectively only change their signs,
\be
A(0)B((2-r)\eta)C(0)D((2-r)\eta) = M_{1}M_{2}M_{3}M_{4}e^{2(2-r)\eta} = \widetilde{V}_{1}^{2}.
\ee
These give the height probability as
\be
P_{a}(b,c,d,e) = \f{\Tr(S_{a}\widetilde{V}_{1}^{2}e^{-2t\eta \mathcal{H}})}{\Tr(\widetilde{V}_{1}^{2}e^{-2t\eta \mathcal{H}})} ,\quad t = \begin{cases}2+r, \quad u\in \mathcal{D}_{1},\\ 2-r, \quad u\in \mathcal{D}_{2}\end{cases}. \label{eq:tDomains}
\ee

The final step in the calculation is determining the diagonal form of the
CTMs. To do this, one first employs the `conjugate modulus transformation', which
gives an expansion of the weights around $p=1$, instead of $p=0$ in the original
formulation of the weights. 
The result of this calculation is that the CTMs are diagonal in the limit $p\rightarrow 1$.
In calculating the matrix elements, the first observation is that the elliptic weights of the model are quasi-periodic in $u$ with the period $2\im K'$, so the elements of $\mathcal{H}$ are integer multiplets of $\pi/K'$,
\be
\mathcal{H}_{{\bl},{\bl'}} = \f{\pi N(\bl)\delta(\bl,\bl')}{K'},
\ee
where $N(\bl)$ is an integer function. In particular, $A$ takes the form $A_{\bl,\bl'}= (e^{-u\mathcal{H}})_{\bl,\bl'} = g_{l_1}^{-1} w^{\phi(\bl)} \delta_{\bl,\bl'}$,
where $\bl = (l_1,\ldots,l_m)$ and $\bl' = (l'_1,\ldots,l'_m)$ label the rows and
columns of $A$; $w = e^{-2\pi \frac{u}{K'}}$, and
$g_{l_1} = w^{\frac{(2l_1-r)^2}{16r}}$. Finally $\phi(\bl)\equiv N(\bl)/2$ is given by
\begin{equation}
\phi(\bl) = \sum_{j=1}^{\frac{m+1}{2}} j \biggl( 
\frac{|l_{2j+3}-l_{2j-1}|}{4} + \delta_{l_{2j-1},l_{2j+1}}\delta_{l_{2j+1},l_{2j+3}}\delta_{l_{2j},l_{2j+2}}
\biggr) \ . \label{eq:phidef}
\end{equation}

Having found the diagonal form of $A$ in the limit $p\rightarrow 1$, one uses the last essential
ingredient of the method, to find the diagonal form for all (positive) $p$.
The function $\phi(\bl)$ takes integer or half-integer values. Because
the weights of the model depend continuously on $p$, it is reasonable to assume that
$\mathcal{H}$ does not change discontinuously with $p$. This in turn implies that
the function $\phi(\bl)$ is in fact independent of $p$, and the diagonal form of $A$ which
was determined for $p=1$, is in fact valid for $0\leq p \leq 1$. With this diagonal form for
$A$, one can give an explicit expression for the local height probabilities $P_a$.

\subsection{The local height probabilities and the function $X_{m}(a;b,c,d,e;q)$}

The local height probabilities can finally be written in the following from

\bea
P_{a}(b,c,d,e) &=& S^{-1}v_{a}X_{m}(a;b,c,d,e; x^{t}).\\
v_{a}&=&x^{(2-t)(2a-r)^{2}/(16r)}E(x^{a},x^{r})\\
S &=& \sum_{a}v_{a}X_{m}(a;b,c,d,e;x^{t})\\
x&=&e^{-4\pi \eta/K'} = e^{-\f{4\pi}{r}K/K'}.
\eea
with boundary conditions $l_{1} = a$ and $l_{m+1}, l_{m+2},\dots = b,c,d,e$ and $m$ is the lattice size.
The variables $p$ and $x$ both lie in the range $0\leq x,p \leq 1$, but when $p\rightarrow 0$, we have $x\rightarrow 1$, and vice versa. The function $E(z,x)$ is the triple product
\be
E(z,x) = \prod_{n=1}^{\infty}(1-x^{n-1}z)(1-x^{n}z^{-1})(1-x^{n}).
\ee
The height probabilities take different forms dependent on the parameter $u$, which enter the expressions
for the probabilities via $t$, as indicated in \eqref{eq:tDomains}. In the case that $u>0$, which we call `regime III', following the notation in
\cite{abf84}, $t=r+2$, i.e. $t$ is greater than zero. In the case $u<0$, `regime II', we have that
$t=2-r$, i.e. $t$ is less than zero. For more details on the regimes, we refer to \cite{abf84,ka12}.

The function $X_{m}(a;b,c,d,e; q)$ is defined as
\begin{gather}
X_{m}(a;b,c,d,e; q) = \sum_{\bl = (a,l_2,\dots,l_{m},b,c,d,e)} q^{\phi(\bl)},
\end{gather}
where $\phi(\bl)$ is the function defined in \eqref{eq:phidef}. The boundary conditions were $l_{1}=a$ and $l_{m+1} = b,\dots, l_{m+4}=e$, and the lattice size $m$ is odd in the composite model. The heights are diagonal in the limit $p\to 1$, and from the definition one can see that
\bea
X_{m}(a;b,c,d,e;q) =q^{\f{m+1}{2}\left(\f{\abs{b-1-e}}{4}+\kd{b-1,c}\kd{c,e}\kd{b,d}\right)} X_{m-2}(a;b-2,b-1,b,c;q) \nonumber\\
\label{eq:xrec}
+q^{\f{m+1}{2}\left(\f{\abs{b-1-e}}{4}+\kd{b-1,c}\kd{c,e}\kd{b,d}\right)}X_{m-2}(a;b,b-1,b,c;q)\\
+q^{\f{m+1}{2}\left(\f{\abs{b+1-e}}{4}+\kd{b+1,c}\kd{c,e}\kd{b,d}\right)}X_{m-2}(a;b,b+1,b,c;q) \nonumber\\
+q^{\f{m+1}{2}\left(\f{\abs{b+1-e}}{4}+\kd{b+1,c}\kd{c,e}\kd{b,d}\right)}X_{m-2}(a;b+2,b+1,b,c;q) \nonumber
\eea
and the states appearing on the right hand side (RHS) only depend on $b,c$. Again, by assuming continuity in $p$, this recursion relation is valid for all $p$. Also, due to the symmetries of the plaquette weights of the model, $X_{m}$ satisfies \cite{abf84,ka12}
\be
X_{m}(r-a;r-b,r-c,r-e;q) = X_{m}(a;b,c,d,e; q).
\label{eq:refl}
\ee

\subsection{Phases of the composite height model}
\label{subsec:phases}

From the expressions of the local height probabilities (or better, the
partition function), we can extract the
phase diagram of the model. Here, we will concentrate on the case
$p\geq 0$, for which the expressions for the LHPs of the previous section are valid.

In this section, we give the ground states in the gapped region $0 < p < 1$.
These ground states will play an important role in making the connection between
the LHP for $p=0$ and the conformal field theory characters.

We start by considering the case $u>0$, i.e. regime III. 
The ground states are those configurations
which contribute maximally to the partition function. As was the case for the LHP, the
dependence on the regime is via the parameter $t$, which is positive in regime III.
This in turn implies that to find the ground state configurations, the function
$\phi(\bl)$ has to be minimized (see \cite{ka12} for more details). The first term in $\phi(\bl)$
vanishes when $l_{2j+3} = l_{2j-1}$, which is a necessary condition in a ground state.
We recall that neighbouring heights have to differ by one.
The first way in which second term in $\phi(\bl)$ also vanishes is when $l_{2j} = l_{2j+2}$ and
$l_{2j+1}= l_{2j-1} \pm 2$. These type of ground state patterns will be denoted by
$G_1^{+}$ (when $l_{2j+1}= l_{2j-1} + 2$) and $G_1^{-}$ (when $l_{2j+1}= l_{2j-1} - 2$).
The second term in $\phi(\bl)$ also vanishes for $l_{2j+1} = l_{2j-1}$ and
$l_{2j} = l_{2j-1}+1 = l_{2j+2}+2$ (these ground states are denoted by $G_2^{+}$), 
or for $l_{2j} = l_{2j-1} - 1 = l_{2j+2} - 2$ (these ground states are denoted by $G_2^{-}$).
This exhausts the possible ground state patterns for $u>0$. These ground state
patterns are depicted, together with the other possible patterns to be discussed
below, in figure \ref{fig:gsfig}.

Turning our attention to the case $u<0$ or regime II, we have that the parameter $t$ is
negative, which implies that the ground states maximize the function $\phi(\bl)$ (see \cite{ka12}).
We start by noting that the first term in the sum in $\phi(\bl)$ can not be one for all
values of $j$, because the heights take their values in the finite range $l=1,2,\ldots r-1$.
The second term in the sum can however always be one, which is thus the case for
the ground states. We find that $l_{2j-1} = l_{2j+1} = l_{2j+3}$ and
$l_{2j} = l_{2j+2} = l_{2j-1} \pm 1$ are the necessary conditions.
These $u<0$ ground state patterns are denoted by  by $G_3^{\pm}$.

Before discussing the transition point $p=0$, we first mention that, if one considers
four consecutive heights, there are only two patterns left, which are not part of a ground
state pattern. These are $(b,b+1,b+1,b+3)$, which we will denote by ${\rm NGS}^{+}$ and
$(b,b-1,b-2,b-3)$, denoted by ${\rm NGS}^{-}$. These patterns will play a role in the study
of the local height probabilities in connection with the critical point at $p=0$.

At the point $p=0$, we find that all configurations contribute to the partition function,
which means that one has to consider the local height probabilities in their entirety.
In \cite{ka12}, it was observed that the local height probabilities are related to (finitized)
characters of certain conformal field theories. This behaviour, as alluded to in the
introduction, has been observed in various other cases as well \cite{saleur88, date86, date87}, not in the least for the
ABF model.

Let us be a bit more precise about the connection between the expressions for the
LHPs and conformal field theory. It turns out that the functions $X_m(a;b,c,d,e;q)$
appearing in the expression for the LHP $P_{a}(b,c,d,e)$ correspond to a character of a conformal field theory, if the boundary
condition $(b,c,d,e)$ is part of a ground state pattern we discussed above. The
conformal field theory is the one describing the critical behaviour at the phase
transition to the phase exhibiting the ground state pattern under consideration.

In particular, in \cite{ka12}, an explicit expression for the function
$X_m(a;b,c,d,e;q)$ for $r=5$ was conjectured, which equals the finitized characters of a
particular CFT. In the following, we prove this conjecture, thereby establishing the
connection between the model for $r=5$, and the CFT.
In \cite{ka12}, the connection with the Gepner parafermions associated with $su(3)_2$ was
made. This parafermionic coset $su(3)_2/(u(1)_{4}\times u(1)_{12})$ is equivalent with the diagonal coset
$su(2)_1 \times su(2)_1 \times su(2)_1 /su(2)_3$, which is only one member of an infinite series
of equivalences, which starts with the equivalence of the $Z_2$ parafermions, $su(2)_2/u(1)_{4}$
and the first minimal model, i.e. the Ising model.
In addition, we provide an explicit
form for the functions $X_m(a;b,c,d,e;q)$ for arbitrary $r$, and prove the result also for
$r=6$. We argue that these functions are the finitized characters of a set of coset
models, similar to the `minimal models' describing the $u>0$ critical point of the
ABF model.

Before we start the discussion of the general properties of the functions $X_m(a;b,c,d,e;q)$
in the next subsection, we note that the number of independent height probabilities,
given the boundary conditions which correspond to ground state patterns, is
$(r-1)(r-3)$ in the regime $u>0$, and $(r-1)(r-2)/2$ in the regime $u<0$. Here, the
reflection symmetry \eqref{eq:refl} was already taken into account to reduce the
number of independent functions $X_m (a;b,c,d,e;q)$.

\subsection{Recursion relations for $X_{m}(a;b,c,d,e;q)$}
\label{sec:Xproperties}
\label{sec:Xrecursion}

We continue by describing some general properties of the functions
$X_m (a;b,c,d,e;q)$, before we deal more explicitly with the cases $r=5$
and $r=6$ in the following sections, where we give an explicit expression
for these functions, and prove that they are equivalent to the functions
$X_m$, by showing that they obey the same recursion relations, and have
identical boundary conditions.

\begin{figure}[t]
\includegraphics[width=.45\textwidth]{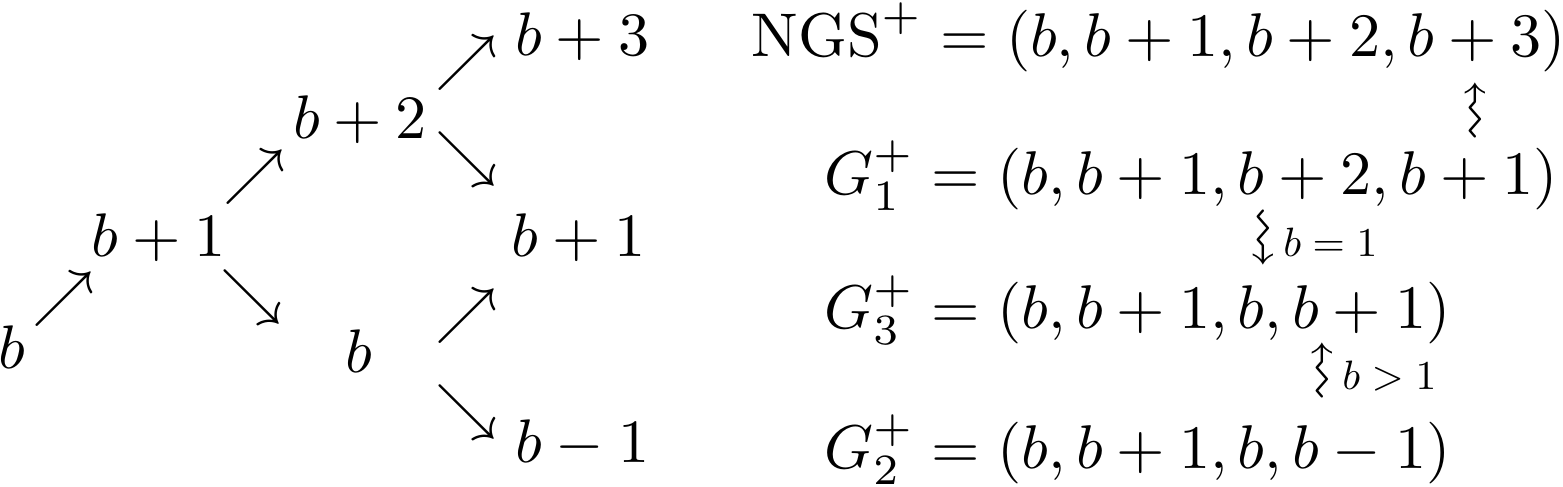}
\hspace{.05\textwidth}
\includegraphics[width=.45\textwidth]{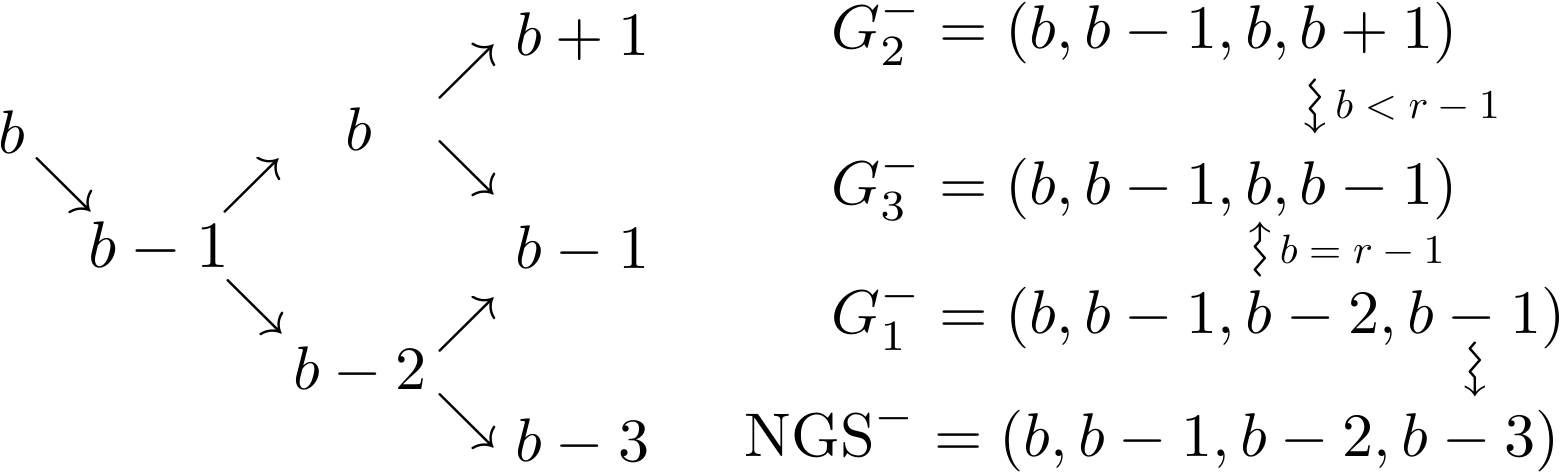}
\caption{The various height configurations that appear in the model. The patterns $G_1^{\pm}$ and $G_2^{\pm}$ are ground states for $u>0$ and the patterns $G_3^{\pm}$ for $u<0$. The patterns $\rm NGS^{\pm}$ are not ground state patterns but appear in the recursion of $X_{m}$. The meaning of the arrows $\leadsto$ is explained in the main text.}
\label{fig:gsfig}
\end{figure}

The possible boundary conditions $(b,c,d,e)$ for the LHP are only constrained
by the fact that neighbouring heights have to differ by one, and are all given
in figure~\ref{fig:gsfig}.

The different states labelled $G_1^{\pm}, G_2^{\pm}, G_3^{\pm}$ and $\rm NGS^{\pm}$ appear
in the recursion for $X_{m}$ as described below.

Although in establishing the connection between the LHPs and conformal field theory, we
are mainly interested in the boundary conditions corresponding to ground states, the
recursion relations force us to consider the non ground state patterns as well, because they
are generated by the recursion relations automatically.

Due to the relation \eqref{eq:refl}, we need to consider only half of the patterns in figure~\ref{fig:gsfig}. We will focus on the patterns with an increasing second height. Starting from an $G_1^{+}$ boundary condition, relevant for $u>0$, one finds
\begin{gather}
X_{m}(a;b,b+1,b+2,b+1;q) =  
q^{\f{m+1}{4}}\underbrace{X_{m-2}(a;b-2,b-1,b,b+1;q)}_{\textrm{NGS}^{+}}
+q^{\f{m+1}{4}}\underbrace{X_{m-2}(a;b,b-1,b,b+1;q)}_{G_2^{-}} \\
+ \underbrace{X_{m-2}(a;b,b+1,b,b+1;q)}_{G_3^{+}}+
\underbrace{X_{m-2}(a;b+2,b+1,b,b+1;q)}_{G_1^{-}}. \nonumber
\end{gather}
Here the first term vanishes for $b\leq 2$ and the second in the case $b=1$.
The third term is a ground state pattern for $u<0$ and the first term is a non-ground state pattern. 

For the ground states in $G_2^{+}$, we get
\begin{gather}
X_{m}(a;b,b+1,b,b-1;q) = 
\underbrace{X_{m-2}(a;b-2,b-1,b,b+1;q)}_{\textrm{NGS}^{+}}
+ \underbrace{X_{m-2}(a;b,b-1,b,b+1;q)}_{G_2^{-}} \\
+ q^{\f{m+1}{4}}\underbrace{X_{m-2}(a;b,b+1,b,b+1;q)}_{G_3^{+}}
+ q^{\f{m+1}{4}}\underbrace{X_{m-2}(a;b+2,b+1,b,b+1;q)}_{G_1^{-}}. \nonumber
\end{gather}
Again, some terms do not necessarily contribute and the first term is not a ground state pattern.

The set of patterns relevant for the $u<0$ LHP lead to a recursion of the form
\begin{gather}
X_{m}(a;b,b+1,b,b+1;q) =
q^{\f{m+1}{4}}\underbrace{X_{m-2}(a;b-2,b-1,b,b+1;q)}_{\textrm{NGS}^{+}} 
+ q^{\f{m+1}{4}}\underbrace{X_{m-2}(a;b,b-1,b,b+1;q)}_{G_2^{-}} \\
+ q^{\f{m+1}{2}}\underbrace{X_{m-2}(a,b,b+1,b,b+1;q)}_{G_3^{+}}
+ q^{\f{m+1}{2}}\underbrace{X_{m-2}(a;b+2,b+1,b,b+1;q)}_{G_1^{-}} \nonumber.
\end{gather}

Finally, the non-ground state pattern NGS$^{+}$ satisfies the the recursion
\begin{gather}
X_{m}(a; b,b+1,b+2,b+3;q) =
q^{\f{m+1}{2}}\underbrace{X_{m-2}(a;b-2,b-1,b,b+1;q)}_{\textrm{NGS}^{+}}
+ q^{\f{m+1}{2}}\underbrace{X_{m-2}(a;b,b-1,b,b+1;q)}_{G_2^{-}}\\
+ q^{\f{m+1}{4}}\underbrace{X_{m-2}(a;b,b+1,b,b+1;q)}_{G_3^{+}}
+ q^{\f{m+1}{4}}\underbrace{X_{m-2}(a;b+2,b+1,b,b+1;q)}_{G_1^{-}} \nonumber.
\end{gather}

It is clear that not only does the recursion relation generate non ground state patterns,
it also mixes the different ground state patterns relevant for the two different regimes
$u>0$ and $u<0$. This implies that we will have to establish the recursion relations for
all types of boundary conditions.

To make our work easier, we start by establishing some relations between the different
functions $X_m(a;b,c,d,e;q)$, which reduces the number of cases we have to check explicitly.
These relations originate in the fact that the states appearing in the recursion only depend
on $b,c$. Using the definition of $X_{m}(a;b,c,d,e;q)$, one can see that changing the boundary height $e\to e\pm 2$ as follows
\begin{eqnarray*}
(b,b+1,b+2,b+1) &\leadsto& (b,b+1,b+2,b+3)\\
(b,b+1,b,b+1) &\leadsto& (b,b+1,b,b-1)
\end{eqnarray*}
leads to the relations  
\bea
\label{xrelngs}
X_{m}(a;b,b+1,b+2,b+3;q) &=& q^{\f{m+1}{4}}X_{m}(a;b,b+1,b+2,b+1;q) \\
\label{xreluneg1}
X_{m}(a;b,b+1,b,b+1;q) &=& q^{\f{m+1}{4}} X_{m}(a;b,b+1,b,b-1;q) \ ,
\eea
where the second equation is only valid for $b>1$. The first equation relates the
the non ground state pattern ${\rm NGS}^{+}$ to the $u>0$ ground state pattern
$G_1^{+}$. The second equation relates the $u<0$ ground state pattern $G_3^{+}$ to the
$u>0$ ground state pattern $G_2^{+}$, for $b>1$. The case $b=1$ can be dealt with
by relating the pattern to the $u>0$ pattern $G_1^{+}$ by changing the boundary height $d$ instead, namely
$(b,b+1,b+2,b+1) \leadsto (b,b+1,b,b+1)$, as follows, (see fig. \ref{fig:gsfig})
\be
X_{m}(a;1,2,1,2;q) = q^{\f{m+1}{2}}X_{m}(a;1,2,3,2;q) .
\label{xreluneg2}
\ee
We note that this relation only holds for $b=1$, and can not be used to reduce the
number of independent functions $X_{m}(a;b,c,d,e;q)$ even further.
In conclusion, we find that all the functions $X_{m}(a;b,c,d,e;q)$ corresponding
to non ground state pattens and $u<0$ ground state patterns can be related to
$u>0$ ground state patters, and we are thus left with $(r-1)(r-3)$ independent
functions, corresponding to, say, the $G_1^{+}$ and $G_2^{-}$ patterns.

\section{Explicit expressions for $r=5$}\label{sec:r=5}

We start our search for explicit expressions for the functions $X_{m} (a;b,c,d,e;q)$
with the case $r=5$. For this case, an explicit functional expression was obtained in \cite{ka12} based on numerical evidence, but
the equivalence was not proven. We provide the proof in this section.

\subsection{The function $y(k;l_{2},l_{3},l_{4};q)$ for $r=5$} 

For $r=5$, the functions $X_{m}(a;b,c,d,e)$ for the different boundary conditions are related to the functions \cite{ka12}
\be
y(k;l_{2},l_{3},l_{4};q) =
\sideset{}{'}\sum_{m_{1},m_{2} \geq 0}
q^{\f{1}{2}(m_{1}^{2}+m_{2}^{2}-m_{1}m_{2}-m_{1} \kd{l_{4},3}- m_{2} \kd{l_{4},2})}
\qbinom{\f{k+m_{2}+\kd{l_{3},1}+\kd{l_{4},3}}{2}}{m_{1}}
\qbinom{\f{k+m_{1}+\kd{l_2,1}+\kd{l_3,2}+\kd{l_4,2}}{2}}{m_{2}}
\label{eq:ydefr5}
\ee
where $l_{4}=1,\dots,4$, $l_{2},l_{3}=1,2$ and
$\qbinom{m}{n} \equiv \qbinom{m}{n}_{q}$ is the $q$-binomial coefficient
\cite{andrews}. First, we define $(q)_{m} = \prod_{j=1}^{m} (1-q^{j})$ for
integer $m>0$ and $(q)_{0} =1$. We then define the $q$-binomials, non-zero for integer $m,n$, as
\begin{equation}
\qbinom{m}{n}_{q} = \begin{cases}
\frac{(q)_{m}}{(q)_{n}(q)_{m-n}} & \text{if $0\leq n \leq m$ integers,} \\
0 & \text{otherwise}
\end{cases} \ .
\end{equation}
The prime on the summation in \eqref{eq:ydefr5} indicates constraints on the parities of $m_{1}, m_{2}$ but with the above definition of the $q$-binomials, they are in fact superfluous and will be therefore left implicit. This, however, will not be the case for the constraints for $r>5$. 

The relations between the $X_{m}$ and $y(k)$ are schematically \cite{ka12}
\be
X_{m}(a;b,c,d,e;q) \sim y(\f{m-1}{2},l_{2},l_{3},l_{4};q),
\ee
where the odd integer $m$ is the lattice size of the composite height model and $l_{2},l_{3},l_{4}$ are determined by the configurations $(a;b,c,d,e)$.

Given this correspondence, the recursion for $X_{m}(a;b,c,d,e;q)$
implies a recursion schematically of the form
\be
y(\f{m-1}{2};l_{2},l_{3},l_{4};q) \sim q^{(m+1)/4}y(\f{m-1}{2}-1;l'_{2},l'_{3},l'_{4};q)
+\cdots+y(\f{m-1}{2}-1;l''_{2},l''_{3},l''_{4};q)+\cdots,
\ee 
or more conveniently in terms of the integer $k=\f{m-1}{2}$,
\be
y(k;l_{2},l_{3},l_{4};q)\sim q^{\f{k+1}{2}}y(k-1;l'_{2},l'_{3},l'_{4};q)
+\cdots+y(k-1;l''_{2},l''_{3},l''_{4};q)+\cdots
\ee
for the functions $y(k;l_{2},l_{3},l_{4};q)$. From now on, we will display the dependence on
the size of the lattice through the (integer valued) variable $k$ instead of the odd integers $m$.

As in the original paper of ABF, we now set out to prove that the functions
$X_{2k+1}(a;b,c,d,e;q)$ and $y(k;l_{2},l_{3},l_{4};q)$ satisfy the same recursion relations and have identical boundary conditions and thus have to agree identically. This verifies the critical properties of the anyon model, since the functions $y(k;l_{2},l_{3},l_{4};q)$ are --- conjecturally for general $r$ --- finitized forms of CFT characters of the coset $\f{su(2)_{1}\times su(2)_{1}\times su(2)_{r-4}}{su(2)_{r-2}}$, as in the original case of RSOS height probabilities and minimal models $\mathcal{M}(r-1,r)$ studied by ABF and by many authors in subsequent papers cited in the introduction. However, due to the composite nature of the height model, the corresponding recursions are more complicatedly related, as we will see. Unfortunately, we have not been able to obtain a functional form that would directly satisfy the recursion of $X_{2k+1}$ and need to proceed in a more oblique way in terms of the more general functions $y(k_{1},k_{2};l_{2},l_{3},l_{4};q)$, presented in the next subsection. The reason behind this
is that the recursion for $X_{2k+1}(a;b,c,d,e;q)$ gives a sum in terms of $X_{2k-1}$, but all with different boundary conditions, while the recursion for $y(k_1,k_2;l_2,l_3,l_4;q)$ leads
to a sum of functions with the same values of the $l_2,\ l_3, \ l_4$, but with different values
for $k_1, \  k_2$.

We first deal with the simplest case $r=5$, corresponding to the diagonal coset
$su(2)_1 \times su(2)_1 \times su(2)_1 / su(2)_3$ (or, equivalently, the 
$su(3)_{2}/(u(1)_{4}\times u(1)_{12})$ Gepner parafermions) in the regime $u>0$ and $Z_{3}$ parafermions for $u<0$, as initiated in the paper \cite{ka12}. In section \ref{sec:r=6}, we prove the correspondence for $r=6$ and give the general conjecture for
arbitrary $r$ in section \ref{sec:conjecture}.

\subsubsection{Recursion for $y(k;l_{2},l_{3},l_{4};q)$}

The function $y(k;l_{2},l_{3},l_{4};q)$ satisfies a recursion relation in $k$,
based on the recursion for $q$-binomial coefficients which follows directly from
the definition \cite{andrews} 
\be
\qbinom{m}{n} = q^{n}\qbinom{m-1}{n}+\qbinom{m-1}{n-1} 
= \qbinom{m-1}{n}+q^{m-n}\qbinom{m-1}{n-1}, \qquad \text{for  $m\geq n \geq 1$} \ .
\label{eq:binomrec}
\ee
In order to have a closed recursion for $y(k;l_{2},l_{3},l_{4};q)$, we define a closely related function, which we will also denote by $y$ and hope that no confusion arises,
\be
\label{eq:ydef}
y(k_{1},k_{2};l_{2},l_{3},l_{4};q) =
\sideset{}{'}\sum_{m_{1},m_{2} \geq 0}
q^{\f{1}{2}(m_{1}^{2}+m_{2}^{2}-m_{1}m_{2}-m_{1} \kd{l_{4},3}- m_{2} \kd{l_{4},2})}
\qbinom{\f{k_{1}+m_{2}+\kd{l_{3},1}+\kd{l_{4},3}}{2}}{m_{1}}
\qbinom{\f{k_{2}+m_{1}+\kd{l_2,1}+\kd{l_3,2}+\kd{l_4,2}}{2}}{m_{2}}
\ee
Clearly $y(k;l_{2},l_{3},l_{4};q) = y(k,k;l_{2},l_{3},l_{4};q)$. 

Then, using the latter recursion in \eqref{eq:binomrec} for $k_{1}$ or $k_{2}$, leads to
\begin{eqnarray}
y(k_{1},k_{2};l_{2},l_{3},l_{4};q) &=&
q^{\f{k_{1}+\kd{l_{3},1}-1}{2}}y(k_{1}-2,k_{2}+1;l_{2},l_{3},l_{4};q) + 
y(k_{1}-2,k_{2};l_{2},l_{3},l_{4};q). \label{eq:m1recursion}\\
&=& q^{\f{k_{2}+\kd{l_{3},2}+\kd{l_{2},1}-1}{2}}y(k_{1}+1,k_{2}-2;l_{2},l_{3},l_{4};q) +
y(k_{1},k_{2}-2;l_{2},l_{3},l_{4};q). \label{eq:m2recursion}
\end{eqnarray}

This is very similar to the recursion for a closely related function function $Y(k_{1},k_{2};q)$
studied in Ref. \cite{arr01}.
Similarly, we can use the recursion in both $k_{1}$ and $k_{2}$ to arrive at
\begin{gather}
y(k_{1},k_{2};l_{2},l_{3},l_{4};q) =
q^{\f{k_{1}+k_{2}+\kd{l_{2},1}}{2}}y(k_{1}-1,k_{2}-1,l_{2},l_{3},l_{4};q)
+y(k_{1}-2,k_{2}-2,l_{2},l_{3},l_{4};q)
\nonumber\\
+q^{\f{k_{1}+\kd{l_{2},1}+\kd{l_{3},2}-1}{2}}y(k_{1}-2,k_{2}-1,l_{2},l_{3},l_{4};q)+q^{\f{k_{2}+\kd{l_{3},1}-1}{2}}y(k_{1}-1,k_{2}-2,l_{2},l_{3},l_{4};q) \ . \label{eq:double_recursion}
\end{gather}

But, as explained above, the recursion with fixed indices $l_{2}, l_{3}, l_{4}$ is not really enough since in the recursion for $X_{2k+1}$, the boundary conditions will change, so the values for $l_{2},l_{3},l_{4}$ will change correspondingly.
We therefore derive a set of relations for the functions $y(k_1,k_2;l_2,l_3,l_4;q)$, which allow us
to change the values of the $l_i$.

\subsubsection{Identities for $y(k_{1},k_{2};l_{2},l_{3},l_{4};q)$}
\label{sec:r5relations}

To derive the necessary identities, we start from the explicit definition of 
$y(k_1,k_2;l_2,l_3,l_4;q)$ in Eq. \eqref{eq:ydef}. The variables $l_2$ and $l_3$ (both taking
the values $l_2,l_3 = 1,2$) only appear in the $q$-binomials. The same is true for $l_4$ if it
takes the values $l_4=1,4$. We can therefore relate the functions $y(k_1,k_2)$ for the two different
values of $l_2$ (keeping $l_3,l_4$ fixed), by shifting the values of $k_1,k_2$, and similarly
for $l_{3}$ (with $l_{2},l_{4}$ fixed).
To relate the functions with the values $l_4=2,3$, we need to swap the values of $k_1$ and
$k_2$ and shift them, where the shifts depend on the values of $l_2,l_3$. 
In particular, we find (suppressing the variable $q$, as we will frequently do as well below)
\begin{equation}
\label{shifts}
\begin{split}
y(k_1,k_2;1,l_3,l_4) &= y(k_1,k_2+1;2,l_3,l_4) \\ 
y(k_1,k_2;l_2,1,l_4) &= y(k_1+1,k_2-1;l_2,2,l_4) \\
y(k_1,k_2;l_2,l_3,1) &= y(k_1,k_2;l_2,l_3,4) \\ 
y(k_1,k_2;1,1,2) &= y(k_2,k_1;1,1,3) \\ 
y(k_1,k_2;1,2,2) &= y(k_2+2,k_1-2;1,2,3) \\ 
y(k_1,k_2;2,1,2) &= y(k_2-1,k_1+1;2,1,3) \\ 
y(k_1,k_2;2,2,2) &= y(k_2+1,k_1-1;2,2,3) \ .
\end{split}
\end{equation}
There are no further relations needed with $l_{4}$ since $l_{4} = a$ (see Sec. \ref{sec:conj_fermionic_form}) and the reflection $a\to r-a$ is the only change possible we can make in the recursion for $X_{2k+1}$. 

In view of the relations \kaava{shifts}, there are two independent functions, say,
\be
y(k_{1},k_{2};1,1,1) \textrm{ and } y(k_{1},k_{2};1,1,2), \nonumber
\ee
which in particular satisfy the following identities
\begin{eqnarray*}
\label{eq:resum}
y(k_{1},k_{2};1,1,1) &=& y(k_{2},k_{1};1,1,1)\\
y(k_{1},k_{2};1,1,2) &=& y(k_{2},k_{1};1,1,3) . 
\end{eqnarray*}

In the following, we show that in fact $y(k_{1},k_{2};1,1,2) = y(k_{1},k_{2};1,1,3)$
identically, without swapping the arguments $k_{1},k_{2}$. From the point of view of the
definition, Eq. \eqref{eq:ydef}, this is a rather nontrivial equation, because one can not
simply relate terms in the sum of $y(k_{1},k_{2};1,1,2)$ to terms in the sum of
$y(k_{1},k_{2};1,1,3)$; all the products of $q$-binomials get mixed. This is in contrast to the
identities in Eq.~\eqref{shifts}, which could be obtained by trivial relabelings.
Below, we derive similar equations for the other possible values of $l_2$ and $l_3$. 

The recursion for $y(k_{1},k_{2};l_{2},l_{3},l_{4})$ does not involve the variable $l_{4}$.
Using the recursion, it is easy to see that the initial conditions
$y(0,0;l_{2},l_{3},l_{4})$, $y(0,1;l_{2},l_{3},l_{4})$, $y(1,0;l_{2},l_{3},l_{4})$
and $y(1,1;l_{2},l_{3},l_{4})$ specify the values of the function
$y(k_{1},k_{2}; l_{2},l_{3},l_{4})$ uniquely. In fact, one can see that
\begin{center}
\begin{tabular}{r  c | c | c | c}
$(l_2,l_3)=$ & $(1,1)$ & $(1,2)$ & $(2,1)$ & $(2,2)$ \\
\hline
$y(0,0;l_2,l_3,2) = y(0,0,l_2,l_3,3)=$ & $1$ & $q^{1/2}$ & $1$ & $1$ \\
$y(1,0;l_2,l_3,2) = y(1,0,l_2,l_3,3)=$ & $1+q$ & $1+q$ & $q^{1/2}$ & $1$ \\
$y(0,1;l_2,l_3,2) = y(0,1,l_2,l_3,3)=$ & $1+q$ & $1+q$ & $1$ & $q^{1/2}$ \\
$y(1,1;l_2,l_3,2) = y(1,1,l_2,l_3,3)=$ & $2q^{1/2}+q^{3/2}$ & $1+q+q^2$ & $1+q$ & $1+q$ \ .
\end{tabular}
\end{center}
This shows that the the functions $y(k_{1},k_{2};l_2,l_3,2)$ and $y(k_{1},k_{2};l_2,l_3,3)$
are in fact identical.

Similar relations for $l_{4}=1, 4$ are trivially true by changing the symmetric summation variables. So we see that, for $r=5$,
\be
y(k_{1},k_{2}; l_{2},l_{3}, l_{4}) = y(k_{1},k_{2}; l_{2},l_{3},r-l_{4}).
\label{eq:l4swap}
\ee
These relations are analogous to the identity \kaava{eq:refl} for the height probability
$X_{2k+1}(a;b,c,d,e)$.

Using these with the last four relations in \kaava{shifts} gives the following, nontrivial
relations
\begin{equation}
\begin{split}
y(k_1,k_2;1,1,2) &= y(k_2,k_1;1,1,2) \\ 
y(k_1,k_2;1,2,2) &= y(k_2+2,k_1-2;1,2,2) \\ 
y(k_1,k_2;2,1,2) &= y(k_2-1,k_1+1;2,1,2) \\ 
y(k_1,k_2;2,2,2) &= y(k_2+1,k_1-1;2,2,2) \ .
\label{eq:shift-identity}
\end{split}
\end{equation}
With help of Eq.~\eqref{eq:l4swap}, these equations also hold for $l_4=3$. In addition, by
making use of the relations in Eq.~\eqref{shifts} and shifting $l_2$ and $l_3$ when necessary, we find that they also hold for $l_4=1$, and hence for $l_4=4$.
Thus, we have
\begin{equation}
\begin{split}
y(k_1,k_2;1,1,l_4) &= y(k_2,k_1;1,1,l_4) \\ 
y(k_1,k_2;1,2,l_4) &= y(k_2+2,k_1-2;1,2,l_4) \\ 
y(k_1,k_2;2,1,l_4) &= y(k_2-1,k_1+1;2,1,l_4) \\ 
y(k_1,k_2;2,2,l_4) &= y(k_2+1,k_1-1;2,2,l_4) \label{eq:r5identities}
\end{split}
\end{equation}
It is important to note that the functional form of the identities we derived does not
depend on the value of $l_4$. This will be very useful in the following, because the
form of the recursion relations for $y(k_1,k_2,l_2,l_3,l_4)$ does not depend on
$l_4$ either. In establishing the connection between $X_{2k+1}(a;b,c,d,e)$ and
$y(k_1,k_2,l_2,l_3,l_4)$, the cases which only differ in the
values of $l_4$ (or $a$, which is the corresponding variable in the functions $X_{2k+1}$)
can be dealt with simultaneously.

\subsection{The identifications between $X_{2k+1}$ and $y(k)$}

We are now ready to state the relations between the functions $X_{2k+1}(a;b,c,d,e;q)$ and $y(k_{1},k_{2};l_{2},l_{3},l_{4};q)$. Note that due to the properties discussed in sec. \ref{sec:Xproperties}, these identifications have slightly different but equivalent form for the states with $u<0$ as compared to \cite{ka12}. Using these identifications, we can show that the recursions for the functions
$y(k_{1},k_{2})$ at special values of the arguments imply those of the functions $X_{2k+1}$ and that the initial conditions agree.  Thus we are able to give the functions $X_{2k+1}$ in a closed form. 

We have two independent ground state patterns $G_1^{+}$ and $G_2^{-}$.
For the ground states $G_2^{-}$ the identifications are
\bea
X_{2k+1}(1;2,1,2,3;q) &=& y(k,k;1,1,1;q)\nonumber\\
X_{2k+1}(3;2,1,2,3;q) &=& y(k,k;1,1,3;q)\nonumber\\
X_{2k+1}(2;3,2,3,4;q) &=& y(k,k;1,2,2;q)\nonumber\\
X_{2k+1}(4;3,2,3,4;q) &=& y(k,k;1,2,4;q).
\eea 
and for the ground states $G_1^{+}$
\bea
X_{2k+1}(2;1,2,3,2;q)&=&y(k,k;2,1,2;q)\nonumber\\
X_{2k+1}(4;1,2,3,2;q)&=&y(k,k;2,1,4;q)\nonumber\\
X_{2k+1}(1;2,3,4,3;q)&=&y(k,k;2,2,1;q)+q^{\f{k+1}{2}}y(k-1,k-1;1,1,1;q)\nonumber\\
X_{2k+1}(3;2,3,4,3;q)&=&y(k,k;2,2,3;q)+q^{\f{k+1}{2}}y(k-1,k-1;1,1,3;q).
\eea
These are all the $(r-1)(r-3) = 8$ independent functions $X_{2k+1}(a;b,c,d,e;q)$ for $r=5$.

From these identifications, all the ground state patterns specified in Ref. \cite{ka12} are obtained using the properties of $X_{2k+1}$'s and $y$'s. In particular, the identities we derived here show that the identification of ground states $G_1^{+}\leadsto G_3^{+}$ is consistent with the somewhat different identification of ground states in terms of the function $y(k_{1},k_{2};l_{2},l_{3},l_{4})$ in Ref. \cite{ka12} for $r=5$, because we always have that $k_{1}=k_{2}=k$ in the expressions for the height probabilities.

We are now ready to show that the identifications made above are
correct. We will first show that assuming that the identification is correct for
$X_{2k-1}$, this implies that the identification is also correct for $X_{2k+1}$. To do
this, we also make use of the recursion relations for $X_{2k+1}(a;b,c,d,e;q)$ and the functions
$y(k_1,k_2;l_2,l_3,l_4;q)$, as well as the various relations between the latter.
We will complete the proof by showing that the initial conditions also agree.

\subsubsection{Recursions for $G_2^{-}$}

To show that the identifications are indeed as given above, we frequently make use
of the various relations between the functions $y(k_{1},k_{2};l_{2},l_{3},l_{4};q)$
as given in section \ref{sec:r5relations}. For clarity, we repeat these relations when we
use them, with one exception. We frequently use Eq.~\eqref{eq:l4swap} to swap
$l_{4} \leftrightarrow 5-l_{4}$, without mentioning this explicitly.

We start by considering the cases  $G_2^{-}$, namely 
the recursion for $X_{2k+1}(1;2,1,2,3)$ implies
\bea
X_{2k+1}(1;2,1,2,3) = q^{\f{k+1}{2}}X_{2k-1}(1;2,1,2,1)+X_{2k-1}(1;2,3,2,1)+X_{2k-1}(1;4,3,2,1)
\eea
or writing the RHS in terms of the functions $y$,
\begin{equation*}
X_{2k+1}(1;2,1,2,3)
= q^{\f{k+1}{2}}q^{k/2}y(k-1,k-1;1,1,1)+y(k-1,k-1;1,2,4)+q^{k/2}y(k-1,k-1;2,1,4) \ .
\end{equation*}
Now, $y(k-1,k-1;1,2,4) = y(k-2,k; 1,1,1)$ and $y(k-1,k-1;2,1,4) = y(k-1,k-2;1,1,1)$, so in total
\ben
X_{2k+1}(1;2,1,2,3)= q^{\f{k+1}{2}}q^{k/2}y(k-1,k-1;1,1,1)+y(k-2,k;1,1,1)
+q^{k/2}y(k-1,k-2;1,1,1) \ .
\een
Using the recursion $y(k+1,k-2;1,1,1) = q^{\f{k+1}{2}}y(k-1,k-1;1,1,1) +y(k-1,k-2;1,1,1)$, we get
\ben
X_{2k+1}(1;2,1,2,3) = q^{k/2}y(k+1,k-2;1,1,1)+y(k-2,k;1,1,1) = y(k,k;1,1,1) \ ,
\een
where we first used the relation $y(k_{1},k_{2};1,1,1)=y(k_{2},k_{1};1,1,1)$, followed by
the recursion relation for $y(k,k;1,1,1)$. We have thus shown that indeed 
$X_{2k+1}(1;2,1,2,3) = y(k,k;1,1,1)$, based on the identification for $2k-1$ and the
recursion relations.
The case $X_{2k+1}(3;2,1,2,3) = y(k,k;1,1,3)$ follows automatically, because all the
relations we used are independent of the actual value for $l_4$ or $a$.

The only $G_2^{-}$ case left to consider is $X_{2k+1} (2;3,2,3,4)$.
The recursion for $X_{2k+1}(2;3,2,3,4)$  is
\bea
X_{2k+1}(2;3,2,3,4) = q^{\f{k+1}{2}}X_{2k-1}(2;1,2,3,2)+q^{\f{k+1}{2}}X_{2k-1}(2;3,2,3,2) + X_{2k-1}(2;3,4,3,2) \ .
\eea
Writing the RHS in terms of $y$'s gives
\bean
X_{2k+1}(2;3,2,3,4) = q^{\f{k+1}{2}}y(k-1,k-1;2,1,2)+q^{\f{k+1}{2}}q^{k/2}y(k-1,k-1;1,2,2)\\
+y(k-1,k-1;1,1,3) \ .
\eean
Transforming everything to equal indices $l_{2},l_{3},l_{4}$, we get
\bean
X_{2k+1}(2;3,2,3,4) = q^{\f{k+1}{2}}\left(y(k,k-3;1,2,2)+q^{k/2}y(k-1,k-1;1,2,2)\right)+y(k,k-2;1,2,2) \ .
\eean
Now, eq. (\ref{eq:shift-identity}) implies
\ben
y(k_{1},k_{2};1,2,2) = y(k_{2}+2,k_{1}-2;1,2,2) \ .
\een
Using this and the recursion gives
\bean
X_{2k+1}(2;3,2,3,4) =q^{\f{k+1}{2}}y(k+1,k-2;1,2,2)+y(k,k-2;1,2,2)
\eean
which again just the recursion for $ y(k,k;1,2,2) = X_{2k+1}(2;3,2,3,4)$. The recursion for $X_{2k+1}(4;3,2,3,4)$ follows similarly independent of $l_{4}=a$, using the identity (\ref{eq:r5identities}).

\subsubsection{Recursions for $G_{1}^{+}$}

For the ground states $G_1^{+}$, the first recursion for $X_{2k+1}(2;1,2,3,2)$ leads to
\bea
X_{2k+1}(2;1,2,3,2) = X_{2k-1}(2;3,2,1,2)+ X_{2k-1}(2;1,2,1,2) \ .
\eea
In terms of the function $y$, the RHS is
\bean
X_{2k+1}(2;1,2,3,2) = y(k-1,k-1;2,2,3)+q^{k/2}y(k-2,k-2;1,1,3) \\
+ q^{k}y(k-1,k-1;2,2,2)
\eean
or
\bean
X_{2k+1}(2;1,2,3,2) = y(k-2,k;2,1,2)+q^{k/2}y(k-2,k-1;2,1,2) + q^{k}y(k-2,k;2,1,2) \ .
\eean
Now \kaava{eq:shift-identity} implies that
\ben
y(k_{1},k_{2};2,1,2) = y(k_{2}-1,k_{1}+1;2,1,2)
\een
and we get, using the recursion once,
\begin{eqnarray}
X_{2k+1}(2;1,2,3,2) = y(k-2,k;2,1,2)+q^{k/2}y(k-2,k+1;2,1,2)
\end{eqnarray}
which is again just the basic recursion. The recursion for $X_{2k+1}(4;1,2,3,2;q)$ follows similarly.

The recursion for $X_{2k+1}(1;2,3,4,3;q)$ leads to
\bea
X_{2k+1}(1;2,3,4,3) = q^{\f{k+1}{2}}X_{2k-1}(1;2,1,2,3)+X_{2k-1}(1;2,3,2,3)+X_{2k-1}(1;4,3,2,3).
\eea
Writing this in terms of $y$'s is
\bean
X_{2k+1}(1;2,3,4,3)  = q^{\f{k+1}{2}}y(k-1,k-1;1,1,1)+q^{k/2}y(k-1,k-1;1,2,1)+y(k-1,k-1;2,1,4).
\eean
This is simply
\bean
X_{2k+1}(1;2,3,4,3) = q^{\f{k+1}{2}}y(k-1,k-1;1,1,1)+q^{k/2}y(k-1,k;2,2,1)+y(k,k-2;2,2,1).
\eean
Now eq. \kaava{eq:r5identities} implies
\ben
y(k_{1},k_{2};2,2,1) = y(k_{2}+1,k_{1}-1;2,2,1),
\een
using this and the recursion gives back $y(k,k;2,2,1) + q^{\f{k+1}{2}}y(k-1,k-1;1,1,1) = X_{2k+1}(1;2,3,4,3)$. The recursion $X_{2k+1}(3;2,3,4,3)$ is identical.

Since $G_1^{+}$ and $G_2^{-}$ give all the independent states and recursions for $X_{2k+1}$, we
have shown that the recursion for $y(k_{1},k_{2};l_{2},l_{3},l_{4})$ implies the recursion for
$X_{2k+1}(a;b,c,d,e)$.

\subsubsection{Initial conditions}

To establish the equality of the functions $X_{2k+1} (a;b,c,d,e)$ and the functions
$y(k,k,l_2,l_3,l_4)$, we still have to verify the initial conditions. We first show that
knowing the functions $y(k_1,k_2;l_2,l_3,l_4)$ for $k_1,k_2 = 0,1$ for all values of
the $l_i$ fixes the functions completely, by means of the recursion relations. We include
the argument in detail here, because we will need it in the discussion of the case
$r=6$.

It follows from the recursion relations that if we know the functions $y$ for all
$(k_1,k_2)$ with $k_1+k_2 \leq n$, where $n \geq 2$, we can construct all the
functions with $k_1+k_2 = n+1$. In the following, we will suppress the dependence
on $l_2,l_3,l_4$. Namely, the function $y(i,n-i+1)$, where
$i\leq n/2$, can be obtained from $y(i,n-i-1)$ and $y(i+1,n-i-1)$, which we both
know by assumption. Similarly, we can obtain $y(n-i+1,i)$ from $y(n-i-1,i)$ and
$y(n-i-1,i+1)$.

From the knowledge of the functions $y(0,0)$, $y(0,1)$, $y(1,0)$ and $y(1,1)$, we
first obtain $y(0,2)$ and $y(2,0)$ from the recursion. The argument above shows
that we now can obtain all the functions. 

It is now a simple matter to check that the initial conditions for the functions $X_{2k+1}(a;b,c,d,e)$,
namely $X_1(a;b,c,d,e)$ and $X_3(a;b,c,d,e)$, indeed correspond to $y(0,0;l_2,l_3,l_4)$ and
$y(1,1;l_2,l_3,l_4)$.

Below, we give these initial conditions for the independent set of functions $X_{2k+1}$, namely
those corresponding to the ground state patterns for $u>0$.
For the ground state patterns
of type $G_2^{-}$, we have
\begin{center}
\begin{tabular}{r | c | c | c | c}
$(a,b)$ & $(1,2)$ & $(2,3)$ & $(3,2)$ & $(4,3)$ \\
\hline
$X_{1} (a;b,b-1,b,b+1)$ &
$q^{1/2}$ & $q^{1/2}$ & $1$ & $1$ \\ 
$X_{3} (a;b,b-1,b,b+1)$ & 
$1+q^{2}$ & $1+q+q^2$ & $2q^{1/2}+q^{3/2}$ & $q^{1/2}+q^{3/2}$ 
\end{tabular}
\end{center}
The results for the ground state patterns of type $G_1^{+}$ read
\begin{center}
\begin{tabular}{r | c | c | c | c}
$(a,b)$ & $(1,2)$ & $(2,1)$ & $(3,2)$ & $(4,1)$ \\
\hline
$X_{1} (a;b,b+1,b+2,b+1)$ &
$q^{1/2}$ & $1$ & $1$ & $0$ \\ 
$X_{3} (a;b,b+1,b+2,b+1)$ & 
$q^{1/2}+q^{3/2}$ & $1+q$ & $1+2q$ & $q^{1/2}$ 
\end{tabular}
\end{center}
This shows that the initial conditions are identical and completes the proof.

\section{Explicit expressions for $r=6$}\label{sec:r=6}

We continue by considering the case $r=6$. The form of the
functions $y^{(A,B,C)}(k_{1},k_{2},k_{3};l_{2},l_{3},l_{4};q)$ will be
motivated in section \ref{sec:conjecture}, where we give a conjectural form
for the functions $y$ for general $r$.

\subsection{The function $y^{(A,B,C)}(k_{1},k_{2},k_{3};l_{2},l_{3},l_{4};q)$ for $r=6$}

For $r=6$, the `finitized CFT characters' related to the functions
$X_{2k+1}(a;b,c,d,e;q)$ take the form
\bea
y^{(A,B,C)}(k_{1},k_{2},k_{3};l_{2},l_{3},l_{4};q) =
\hspace{-1cm}
\sum_{(m_{1},m_{2},m_{3})\in (2\integers_{\geq 0})^3+(A, B, C)\bmod 2}
\hspace{-1cm}
q^{\f{1}{2}(m_{1}^{2}+m_{2}^{2}+m_{3}^{2} -(m_{1} + m_{3}) m_{2}
-m_{1} \kd{l_{4},4} - m_{2} \kd{l_{4},3} - m_{3} \kd{l_{4},2})}  \nonumber\\ 
\times
\qbinom{\f{k_{1} + m_{2} + \kd{l_{3},1} + \kd{l_{4},4}}{2}}{m_{1}}
\qbinom{\f{k_{3} + m_{1} +m_{3} + \kd{l_{3},2} + \kd{l_{4},3}}{2}}{m_{2}}
\qbinom{\f{k_{2}+m_{2}+\kd{l_{2},1}+\kd{l_{3},3}+\kd{l_{4},2}}{2}}{m_{3}},
\label{eq:ydefr6}
\eea
where the summations are restricted such that the $m_i$ have the same parities
as $(A,B,C)$, which now have to be explicitly specified in contrast to the case $r=5$.
We will only consider these functions for $(l_2+l_3+l_4) \bmod 2 = 1$. The properties
of this function derived below are valid under this condition.

We start by noting that the condition $(l_2+l_3+l_4) \bmod 2 = 1$ implies that for
the function $y_{r=6}$ to be non-zero, one needs that $k_{1} = k_{2} \bmod 2$.
This follows from the requirement that the arguments of the $q$-binomials have
to be integers. Thus, one requires that both 
$B = (k_{1} + \kd{l_{3},1} + \kd{l_{4},4}) \bmod 2$ and 
$B = (k_{2} + \kd{l_{2},1} + \kd{l_{3},3} + \kd{l_{4},2}) \bmod 2$.
Inspection shows that to satisfy both equations, one needs $k_{1} = k_{2} \bmod 2$.
In addition, one needs that
$A+C = (k_{3} + \kd{l_{3},2}+\kd{l_{4},3}) \mod 2$, so there are two, a priori independent,
functions for every $l_{2},l_{3},l_{4}$, given by the two different choices for $A$ and $C$.

Again, the height probabilities $X_{2k+1}$ are related to these more general functions with
$k_{1} = k_{2}$, $k_{3} = 0$. The variable $k_{3}$ is introduced in
$y^{(A,B,C)}(k_{1},k_{2},k_{3})$ to obtain a recursion that closes. When we state the
relation between the functions $X_{2k+1} (a;b,c,d,e;q)$ and
$y^{(A,B,C)} (k,k,0;l_2,l_3,l_4;q)$ below, we will specify the required values of
$(A,B,C)$ explicitly.

\subsubsection{Recursion for $y^{(ABC)}(k_{1},k_{2},k_{3};l_{2},l_{3},l_{4};q)$}
The recursion for this function, following from \eqref{eq:binomrec} is, with
$\ell=l_{2},l_{3},l_{4}$ to shorten the notation, 
\bea
y^{(A,B,C)}(k_{1},k_{2},k_{3};\ell;q) 
&=&q^{\f{k_{1}+\kd{l_{3},1}-1}{2}} y^{(A+1,B,C)}(k_{1}-2,k_{2},k_{3}+1;\ell;q)+y^{(A,B,C)}(k_{1}-2,k_{2},k_{3};\ell;q) \nonumber\\
&=&q^{\f{k_{2}+\kd{l_{3},3}+\kd{l_{2},1}-1}{2}}y^{(A,B,C+1)}(k_{1},k_{2}-2,k_{3}+1;\ell;q)+y^{(A,B,C)}(k_{1},k_{2}-2,k_{3};\ell;q) \label{eq:yr=6} \\
&=&q^{\f{k_{3}+\kd{l_{3},2}-1}{2}} y^{(A,B+1,C)}(k_{1}+1,k_{2}+1,k_{3}-2;\ell;q)+y^{(A,B,C)}(k_{1},k_{2},k_{3}-2;\ell;q) \nonumber.
\eea

As was the case for $r=5$, we need to establish several identities for the functions
$y^{(A,B,C)} (k_{1},k_{2},k_{3};l_{2},l_{3},l_{4};q)$, which allow us
to change the values of $l_2,l_3,l_4$, in order to make connection with the recursion for
$X_{2k+1}$, as given in equation \eqref{eq:xrec}, which does not keep the $l_{i}$ constant
throughout the recursion.

\subsubsection{Identities for $y^{(A,B,C)}(k_{1},k_{2},k_{3};l_{2},l_{3},l_{4};q)$}

The relations we need in order to prove the equivalence between the $X_{2k+1}$ and
$y^{(A,B,C)} (k_1,k_2,k_3)$ can be derived in the same way as for $r=5$. However, we need to
keep track of the parities $(A,B,C)$, complicating matters slightly. As explained above,
we only consider the cases obeying $(l_{2} + l_{3} + l_{4}) \bmod 2 = 1$.

Similar to the situation for $r=5$, the relations can be grouped into several
classes. We first deal with the relations which can be obtained from the definition
\eqref{eq:ydefr6} by shifting the values of the $k_i$, and if necessary changing
the summation variables $m_{1} \leftrightarrow m_{3}$. All these relations are
therefore trivial in nature.

The first class of relations relates the functions which only differ in the values of
$l_2$ and $l_3$. For $l_4$ odd, we have (dropping the $q$ dependence)
\begin{equation}
y^{(A,B,C)} (k_{1},k_{2},k_{3};1,1,l_{4}) = 
y^{(A,B,C)} (k_{1}+1,k_{2}+1,k_{3}-1;2,2,l_{4}) = 
y^{(A,B,C)} (k_{1}+1,k_{2}-1,k_{3};1,3,l_{4}) \ .
\end{equation}
In the case $l_4$ even, we find
\begin{equation}
y^{(A,B,C)} (k_{1},k_{2},k_{3};1,2,l_{4}) = 
y^{(A,B,C)} (k_{1}-1,k_{2}+1,k_{3}+1;2,1,l_{4}) = 
y^{(A,B,C)} (k_{1},k_{2},k_{3}+1;2,3,l_{4}) \ .
\end{equation}

We continue by relating the functions $y$ with $l_4$ and $r-l_4$, by
changing the summation variables $m_{1} \leftrightarrow m_{3}$, which
also swaps the values of $k_{1}$ and $k_{2}$. For $l_4$ odd, this gives
\begin{align}
y^{(A,B,C)} (k_{1},k_{2},k_{3},1,1,l_4) &=
y^{(C,B,A)} (k_{2},k_{1},k_{3},1,1,r-l_4) \nonumber \\ 
y^{(A,B,C)} (k_{1},k_{2},k_{3},2,2,l_4) &=
y^{(C,B,A)} (k_{2},k_{1},k_{3},2,2,r-l_4) \nonumber \\ 
y^{(A,B,C)} (k_{1},k_{2},k_{3},1,3,l_4) &=
y^{(C,B,A)} (k_{2}+2,k_{1}-2,k_{3},1,3,r-l_4) \ . \label{eq:r6resum}
\end{align}
For $l_4$ even, we find
\begin{align}
y^{(A,B,C)} (k_{1},k_{2},k_{3},1,2,l_4) &=
y^{(C,B,A)} (k_{2}+1,k_{1}-1,k_{3},1,2,r-l_4) \nonumber \\ 
y^{(A,B,C)} (k_{1},k_{2},k_{3},2,1,l_4) &=
y^{(C,B,A)} (k_{2}-1,k_{1}+1,k_{3},2,1,r-l_4) \nonumber \\ 
y^{(A,B,C)} (k_{1},k_{2},k_{3},2,3,l_4) &=
y^{(C,B,A)} (k_{2}+1,k_{1}-1,k_{3},2,3,r-l_4) \ .
\end{align}
We now relate the functions $y^{(A,B,C)} (k_1,k_2,k_3;l_2,l_3,l_4)$ with $l_4$ and $r-l_4$, but
without changing the summation variables $m_{1} \leftrightarrow m_{3}$.
This can be done trivially in the case $l_4=1,5$, but in the other cases, the
relations are nontrivial, because they completely scramble the contributions
from the products of the binomials, and hence can not be obtained by reshuffling the
terms in the sums in equation \eqref{eq:ydefr6}. Instead, these relations are obtained
by using that the recursion relations for $y$ are independent of $l_4$, and
checking the initial conditions. The initial conditions sometimes give rise to constraints
for the $k_{i}$. We always assume that $k_{1},k_{2} \geq 0$, and give the constraint on
$k_{3}$ explicitly.
\begin{align}
y^{(A,B,C)} (k_{1},k_{2},k_{3},l_{2},l_{3},1) &=
y^{(A,B,C)} (k_{1},k_{2},k_{3},l_{2},l_{3},5) \nonumber \\
y^{(A,B,C)} (k_{1},k_{2},k_{3},l_{2},l_{3},3) &=
y^{(A+1,B,C+1)} (k_{1},k_{2},k_{3},l_{2},l_{3},3)
&& \text{for $k_{3} + \kd{l_{3},2} \geq 0$}\nonumber \\
y^{(A,B,C)} (k_{1},k_{2},k_{3},l_{2},l_{3},2) &=
y^{(A+1,B+1,C+1)} (k_{1},k_{2},k_{3},l_{2},l_{3},4)
&& \text{for $k_{3} + \kd{l_{3},2} \geq 0$} \ .
\end{align}
Finally, we combine these identities with the preceding ones, to obtain
expressions relating the functions $y$ with the same values of
$l_{2},l_{3},l_{4}$, but with the values of $k_{1}$ and $k_{2}$ swapped.
In particular, for $l_{4}$ odd we have
\begin{align}
y^{(A,B,C)} (k_{1},k_{2},k_{3},1,1,l_4) &=
y^{(C+\kd{l_4,3},B,A+\kd{l_4,3})} (k_{2},k_{1},k_{3},1,1,l_4)
&& \text{for $k_{3} \geq 0$} \nonumber \\ 
y^{(A,B,C)} (k_{1},k_{2},k_{3},2,2,l_4) &=
y^{(C+\kd{l_4,3},B,A+\kd{l_4,3})} (k_{2},k_{1},k_{3},2,2,l_4) 
&& \text{for $k_{3} \geq -1$} \nonumber \\ 
y^{(A,B,C)} (k_{1},k_{2},k_{3},1,3,l_4) &=
y^{(C+\kd{l_4,3},B,A+\kd{l_4,3})} (k_{2}+2,k_{1}-2,k_{3},1,3,l_4)
&& \text{for $k_{3} \geq 0$} \ . \label{eq:133identity}
\end{align}
For $l_4$ even, we finally obtain
\begin{align}
y^{(A,B,C)} (k_{1},k_{2},k_{3},1,2,l_4) &=
y^{(C+1,B+1,A+1)} (k_{2}+1,k_{1}-1,k_{3},1,2,l_4)
&& \text{for $k_{3} \geq -1$} \nonumber \\ 
y^{(A,B,C)} (k_{1},k_{2},k_{3},2,1,l_4) &=
y^{(C+1,B+1,A+1)} (k_{2}-1,k_{1}+1,k_{3},2,1,l_4)
&& \text{for $k_{3} \geq 0$} \nonumber \\
y^{(A,B,C)} (k_{1},k_{2},k_{3},2,3,l_4) &=
y^{(C+1,B+1,A+1)} (k_{2}+1,k_{1}-1,k_{3},2,3,l_4)
&& \text{for $k_{3} \geq 0$} \ \label{eq:l4evenidentity}.
\end{align}
This exhausts the relations that we need to prove the equivalence between the
functions $X_{2k+1}$ and $y^{(A,B,C)}(k_1,k_2,k_3)$.

\subsection{The identifications between $X_{2k+1}$ and $y^{(A,B,C)}(k_1,k_2,k_3)$}
\label{sec:r6ident}

As was the case for $r=5$, it suffices to give the identifications for an independent set
of $X_{2k+1}$. We again specify the cases corresponding to the ground state patterns
$G_2^{-}$ and $G_1^{+}$. We henceforth drop commas from the boundary conditions,
$b,c,d,e \to bcde$, the parities of the summation variables $A,B,C \to ABC$ and the
labels $l_{2},l_{3},l_{4} \to l_{2}l_{3}l_{4}$ to lighten the notation. 

For the patterns in $G_2^{-}$, the identifications are
\bea
X_{2k+1}(1;2123) &=& y^{(AAA)}(k,k,0;111) \nonumber \\
X_{2k+1}(3;2123) &=& y^{(AAC)}(k,k,0;113) \nonumber \\
X_{2k+1}(5;2123) &=& y^{(ABA)}(k,k,0;115) \nonumber \\
X_{2k+1}(2;3234) &=& y^{(ABB)}(k,k,0;122) \nonumber \\
X_{2k+1}(4;3234) &=& y^{(AAC)}(k,k,0;124) \nonumber \\
X_{2k+1}(1;4345) &=& y^{(ABA)}(k,k,0;131) \nonumber \\
X_{2k+1}(3;4345) &=& y^{(ABB)}(k,k,0;133) \nonumber \\
X_{2k+1}(5;4345) &=& y^{(AAA)}(k,k,0;135).
\label{eq:xyr6b}
\eea
In the case of patterns of type $G_1^{+}$, they read
\bea
X_{2k+1}(2;1232) &=&y^{(AAA)}(k,k,0;212) \nonumber \\
X_{2k+1}(4;1232) &=&y^{(ABA)}(k,k,0;214) \nonumber \\
X_{2k+1}(1;2343) &=& y^{(ABB)}(k,k,0;221)+q^{\f{k+1}{2}}y^{(BBB)}(k-1,k-1,0;111) \nonumber \\
X_{2k+1}(3;2343) &=& y^{(ABA)}(k,k,0;223)+q^{\f{k+1}{2}}y^{(BBA)}(k-1,k-1,0;113) \nonumber \\
X_{2k+1}(5;2343) &=& y^{(AAC)}(k,k,0;225)+q^{\f{k+1}{2}}y^{(CAC)}(k-1,k-1,0;115) \nonumber \\
X_{2k+1}(2;3454) &=&y^{(ABA)}(k,k,0;232)+q^{\f{k+1}{2}}y^{(BAA)}(k-1,k-1,0;122)
+q^{\f{2k+1}{2}}y^{(BBB)}(k-1,k-1,0;212) \nonumber \\
X_{2k+1}(4;3454) &=&y^{(AAA)}(k,k,0;234)+q^{\f{k+1}{2}}y^{(BBA)}(k-1,k-1,0;124)
+q^{\f{2k+1}{2}}y^{(BAB)}(k-1,k-1,0;214).
\label{eq:xyr6a}
\eea

We still need to specify the parities $(A,B,C)$ for the functions $y$, in order to completely
determine the identification. These read as follows
\begin{align}
A & = (k + l_3 + \kd{l_{4},5} + \kd{l_{3},2}) \bmod 2 \nonumber \\
B & = (k + l_3 + \kd{l_{4},4} + \kd{l_{3},3}) \bmod 2 \nonumber \\
C & = (k + l_3 + \kd{l_{4},3} + \kd{l_{4},5}) \bmod 2 \ .
\label{eq:parities}
\end{align}
It follows that all the parities are reversed whether considering $k$ even or odd.
The notation for the parities in the equations \eqref{eq:xyr6a} and \eqref{eq:xyr6b}
requires some explanation. Clearly we need at most two different labels for the parities.
The parity $A$ of the first function $y^{(A,B,C)}(k,k,0;l_{2},l_{3},l_{4})$ for size $k$ is used as a
reference, and is always denoted by $A$. This parity is obtained from equation \eqref{eq:parities}.
If the parity for $B$, also given by equation \eqref{eq:parities}, happens to be the same
as $A$, this is also denoted $A$. Otherwise, it is denoted as $B$. Finally, for the parity
$C$, the notation is such that it is denoted by $A$ if $A=B=C$, denoted by $B$ if
$A\neq B = C$, and denoted by $C$ if $A=B \neq C$. When the value of $k$ is lowered,
as occurs for the $G_1^{+}$ patterns, see equation \eqref{eq:xyr6a}, the notation of the
parities is with respect to those of size $k$. This notation is convenient to
keep track of the parities, when we prove (in the next section)
the connection between $X_{2k+1}$ and the $y^{(A,B,C)}(k,k,0)$, as given in \eqref{eq:xyr6a}
and \eqref{eq:xyr6b}, and write the parities with respect to the function for size $k$ as
described above.

The identifications given above deal with all the $(r-1)(r-3) = 15$
independent configurations for the functions $X_{2k+1}$ for $r=6$.
The relations for the functions $X_{2k+1}$ for the ground state patterns related
to $u<0$, and the non ground state patterns, can be obtained by making use of
the relations given in the equations \eqref{xreluneg1}, \eqref{xreluneg2} and
\eqref{xrelngs}.

\subsubsection{Recursion for states $G_1^{+}$}
Here we show how to deal with a representative state in $G_1^{+}$, the rest of the recursions
are collected in appendix~\ref{app:r6recursions}. The strategy is the same as for $r=5$,
namely we assume that the identification is correct for $2k-1$, and show that this
implies the identification for $2k+1$, by using the recursions for $X_{2k+1}$
and $y^{(A,B,C)}(k_1,k_2,k_3)$ as well as the various relations between the functions
$y$.
The recursion for $X_{2k+1}(3;2343)$ is
\bea
X_{2k+1}(3;2343) = q^{\f{k+1}{2}}X_{2k-1}(3;2123) + X_{2k-1}(3;2323) + X_{2k-1}(3;4323)
\eea
In terms of $y^{(ABC)}(k_{1},k_{2},k_{3};l_{2},l_{3},l_{4})$ the RHS is
\bean
X_{2k+1}(3;2343) = q^{\f{k+1}{2}}y^{(BBA)}(k-1,k-1,0;113)+q^{\f{k}{2}}y^{(BAA)}(k-1,k-1,0;133)\\
+y^{(BAB)}(k-1,k-1,0;223)+ q^{\f{k}{2}}y^{(AAB)}(k-2,k-2,0;113).
\eean
We use
\bean
y^{(ABC)}(k_{1},k_{2},k_{3};133) &=& y^{(ABC)}(k_{1},k_{2}+2,k_{3}-1;223),\\
y^{(ABC)}(k_{1},k_{2},k_{3};113) &=& y^{(ABC)}(k_{1}+1,k_{2}+1,k_{3}-1;223),
\eean
except for the first term which is included in $X_{2k+1}(3;2343)$, to get
\bean
X_{2k+1}(3;2343)  = q^{\f{k+1}{2}}y^{(BBA)}(k-1,k-1,0;113) + y^{(BAB)}(k-1,k-1,0;223)\\+q^{\f{k}{2}}y^{(BAA)}(k-1,k+1,-1;223)
+q^{\f{k}{2}}y^{(AAB)}(k-1,k-1,-1;223).
\eean
Where the term the second term on the RHS is
\bean
y^{(BAB)}(k-1,k-1,0;223) = y^{(BAB)}(k-1,k-1,-2;223)+y^{(BBB)}(k,k,-2;223),
\eean
so
\bean
X_{2k+1}(3;2343)  = q^{\f{k+1}{2}}y^{(BBA)}(k-1,k-1,0;113)+ y^{(BAB)}(k-1,k-1,-2;223)+y^{(BBB)}(k,k,-2;223)\\
+q^{\f{k}{2}}y^{(BAA)}(k-1,k+1,-1;223)+q^{\f{k}{2}}y^{(AAB)}(k-1,k-1,-1;223).
\eean
Combining the second and last terms gives
\bean
X_{2k+1}(3;2343) = q^{\f{k+1}{2}}y^{(BBA)}(k-1,k-1,0;113) + y^{(BAB)}(k+1,k-1,-2;223)+y^{(BBB)}(k,k,-2;223)\\
+q^{\f{k}{2}}y^{(BAA)}(k-1,k+1,-1;223)
\eean
and further, using the properties in \eqref{eq:133identity} and \eqref{eq:r6resum},
\bean
y^{(ABC)}(k_{1},k_{2},k_{3};223) = y^{(A+1BC+1)}(k_{1},k_{2},k_{3};223) = y^{(CBA)}(k_{2},k_{1},k_{3};223), \quad k_{3}\geq -1,\\
\eean
in the last term gives
\bean
X_{2k+1}(3;2343) = q^{\f{k+1}{2}}y^{(BBA)}(k-1,k-1,0;113) + y^{(BAB)}(k+1,k-1,-2;223)+y^{(BBB)}(k,k,-2;223)\\
+q^{\f{k}{2}}y^{(BAA)}(k+1,k-1,-1;223).
\eean
This is just
\bean
X_{2k+1}(3;2343)  = q^{\f{k+1}{2}}y^{(BBA)}(k-1,k-1,0;113)+ y^{(BAB)}(k+1,k+1,-2;223)+y^{(BBB)}(k,k,-2;223) \\
= q^{\f{k+1}{2}}y^{(BBA)}(k-1,k-1,0;113) + y^{(BBB)}(k,k,0;223) \\ = q^{\f{k+1}{2}}y^{(BBA)}(k-1,k-1,0;113) + y^{(ABA)}(k,k,0;223),
\eean
as desired. The recursions for the same patterns but different $a=l_{4}$, i.e.
$X_{2k+1}(1;2343)$ and $X_{2k+1}(5;2343)$, are similar and omitted.
The remaining recursions for patterns in $G_1^{+}$ are collected in appendix~\ref{app:r6recursions}.

\subsubsection{Recursion for states $G_2^{-}$}

Here we show an example recursion for a state in $G_2^{-}$. We start with
\bea
X_{2k+1}(3;2123) = q^{\f{k+1}{2}}X_{2k-1}(3;2121)+X_{2k-1}(3;2321)+X_{2k-1}(3;4321). 
\eea
In terms of the functions $y^{(ABC)}(k_{1},k_{2},k_{3};l_{2},l_{3},l_{4})$, the RHS is
\bean
X_{2k+1}(3;2123) = q^{\f{k+1}{2}}q^{\f{k}{2}}y^{(CCA)}(k-1,k-1,0;113)+y^{(CAA)}(k-1,k-1,0;133)\\
+q^{\f{k}{2}}y^{(CAC)}(k-1,k-1,0;223)+q^{k}y^{(AAC)}(k-2,k-2,0;113).
\eean
Using 
\bean
y^{(ABC)}(k_{1},k_{2},k_{3};223) &=& y^{(ABC)}(k_{1}-1,k_{2}-1,k_{3}+1;113),\\
y^{(ABC)}(k_{1},k_{2},k_{3};133) &=& y^{(ABC)}(k_{1}-1,k_{2}+1,k_{3};113),
\eean
this is
\bean
X_{2k+1}(3;2123) = q^{\f{k+1}{2}}q^{\f{k}{2}}y^{(CCA)}(k-1,k-1,0;113)+y^{(CAA)}(k-2,k,0;113)\\
+q^{\f{k}{2}}y^{(CAC)}(k-2,k-2,1;113)+q^{k}y^{(AAC)}(k-2,k-2,0;113).
\eean
We use $y^{(ABC)}(k_{1},k_{2},k_{3};113) = y^{(CBA)}(k_{2},k_{1},k_{3};113)$ from \eqref{eq:r6resum} and combine it with the recursion,
\ben
y^{(AAC)}(k-2,k-2,2;113)=q^{1/2}y^{(CCA)}(k-1,k-1,0;113)+y^{(AAC)}(k-2,k-2,0;113),
\een
and get
\bean
X_{2k+1}(3;2123) = q^{k}y^{(AAC)}(k-2,k-2,2;113)+y^{(CAA)}(k-2,k,0;113)\\
+q^{k/2}y^{(CAC)}(k-2,k-2,1;113).
\eean
Using
\ben
y^{(CAC)}(k,k-2,1;113) = y^{(CAC)}(k-2,k-2,1;113)+q^{\f{k}{2}}y^{(AAC)}y(k-2,k-2,2;113),
\een
we get
\bean
X_{2k+1}(3;2123) =  q^{\f{k}{2}}y^{(CAC)}(k,k-2,1;113)+y^{(CAA)}(k-2,k,0;113) = y^{(AAC)}(k,k,0;113),
\eean
where we used $y^{(CAA)}(k-2,k,0;113) = y^{(AAC)}(k-2,k,0;113)$ (recall that the
parity $A\neq C$) and the recursion for $y$. Thus, the identification
$X_{2k+1}(3;2123) = y^{(AAC)}(k,k,0;113)$ follows from the identifications for
$2k-1$ and the recursion for $y$, as we intended to show.

The recursions for $X_{2k+1}(1;2123)$ and $X_{2k+1}(5;2123)$ with different $a=l_{4}$
are very similar and therefore omitted. The recursions for the other patterns in $G_2^{-}$ are
given in appendix~\ref{app:r6recursions}.

\subsubsection{Initial conditions}
We have now proven the recursions for the independent states in $G_1^{+}$ and $G_2^{-}$ are implied by the recursion for the function $y_{r=6}$ and it remains to check that the initial conditions agree. 

We show that to specify the functions $y_{r=6}$ completely, it suffices to know them for all the
values of $k_i \in \{0,1\}$. The others then follow by making use of the recursion relations.
The argument is very similar to the one we gave for the case $r=5$.

We start by giving the recursion relations in symbolic form,
\begin{eqnarray*}
(k_{1},k_{2},k_{3}) &\to& (k_{1}-2,k_{2},k_{3}) + (k_{1}-2,k_{2},k_{3}+1) \ ,
\end{eqnarray*}
and similarly for $k_{1}\leftrightarrow k_{2}$, which occur symmetrically in the
recursion. The value of $k_{3}$ can be lowered by using 
\begin{eqnarray*}
(k_{1},k_{2},k_{3}) &\to& (k_{1},k_{2},k_{3}-2) + (k_{1}+1,k_{2}+1,k_{3}-2) \ .
\end{eqnarray*}

It is straightforward to show that from the initial values, we can obtain all functions
$y_{r=6}$ where the $k_{i}$ satisfy $k_{1} + k_{2} + k_{3} \leq 3$. We start by
constructing $(k_{1},k_{2},k_{3})=(2,0,0)$, $(0,2,0)$ and $(0,0,2)$, by using the functions
for the values $(0,0,0)$, $(0,0,1)$ and $(1,1,0)$, which we know by assumption. We can also
construct $(3,0,0)$, $(0,3,0)$ and $(0,0,3)$, by using the values for $(1,0,0)$, $(1,0,1)$,
$(0,1,0)$, $(0,1,1)$ and finally $(0,0,1)$, and $(1,1,1)$. The functions for these values
also give us the functions at the values $(2,1,0)$ and $(1,2,0)$.

Above, we constructed $y_{r=6}$ for $(0,0,2)$, which together with $(0,0,1)$ gives
us $y_{r=6}$ for $(2,0,1)$ and $(0,2,1)$. Finally, we construct $y_{r=6}$ for the $k_{i}$
values $(1,0,2)$ and $(0,1,2)$ by using $(1,0,0)$, $(0,1,0)$ and the newly constructed
$(1,2,0)$ and $(2,1,0)$. Thus, we now know $y_{r=6}$ for all $k_{i}$ with
$k_{1} + k_{2} + k_{3} \leq 3$.

We continue by showing that knowing the function $y_{r=6}$ for
$k_{1} + k_{2} + k_{3} \leq n$ allows us to construct the function for all $k_{i}$ values
satisfying $k_{1} + k_{2} + k_{3} = n+1$. We start by using the first recursion, or
its equivalent by swapping $k_{1} \leftrightarrow k_{2}$, to construct $y_{r=6}$
for the values $(i,n-i+1,0)$, by using $(i,n-i-1,0)$ and $(i,n-i-1,1)$ or by using
$(i-2,n-i+1,0)$ and $(i-2,n-i+1,1)$, whichever is applicable. We can similarly
construct $(i,n-i,1)$ by using either $(i,n-i-2,1)$ and $(i,n-i-2,2)$ or
$(i-2,n-i,1)$ and $(i-2,n-i,2)$.

From now on, we can exclusively use the second recursion, to construct the remaining
functions. We start by constructing $(i,n-i-1,2)$ from $(i,n-i-1,0)$ and
$(i+1,n-i,0)$ (which was just constructed above). We then continue to subsequently
construct $(i,n-i-j+1,j)$, for $j=3,\ldots,n+1$, from $(i,n-i-j+1,j-2)$ and
$(i+1,n-i-j+2,j-2)$, which have been constructed earlier. This concludes the proof that
we can construct the function $y_{r=6}$ for all values of $(k_{1},k_{2},k_{3})$ from the
knowledge of the functions for the values $k_{i}\in \{0,1\}$ and the recursion relations.

The initial conditions are now straightforward to check, namely
the functions $y^{(ABC)}(0,0,0;l_{2},l_{3},l_{4})$, $y^{(ABC)}(1,1,0;l_{2},l_{3},l_{4})$ and
$X_{1}(a;b,c,d,e)$, $X_{3}(a;b,c,d,e)$ agree. Explicitly, the initial conditions read

\begin{center}
\begin{tabular}{r | c | c | c | c | c | c | c | c}
$(a,b)$ & $(1,2)$ & $(1,4)$ & $(2,3)$ & $(3,2)$ & $(3,4)$ & $(4,3)$ & $(5,2)$ & $(5,4)$ \\
\hline
$X_{1} (a;b,b-1,b,b+1)$ &
$q^{1/2}$ & $0$ & $q^{1/2}$ & $1$ & $q^{1/2}$ & $1$ & $0$ & $1$ \\ 
$X_{3} (a;b,b-1,b,b+1)$ & 
$1+q^{2}$ & $q^{3/2}$ & $1+q+q^2$ & $2q^{1/2}+q^{3/2}$ & $1+q+q^2$ &
$2q^{1/2}+q^{3/2}$ & $q$ & $q^{1/2}+q^{3/2}$ 
\end{tabular}
\end{center}
for ground state patterns of type $G_2^{-}$, while for the ground state patterns of type
$G_1^{+}$, we have
\begin{center}
\begin{tabular}{r | c | c | c | c | c | c | c }
$(a,b)$ & $(1,2)$ & $(2,1)$ & $(2,3)$ & $(3,2)$ & $(4,1)$ & $(4,3)$ & $(5,2)$ \\
\hline
$X_{1} (a;b,b+1,b+2,b+1)$ &
$q^{1/2}$ & $1$ & $q^{1/2}$ & $1$ & $0$ & $1$ & $0$ \\ 
$X_{3} (a;b,b+1,b+2,b+1)$ & 
$q^{1/2}+q^{3/2}$ & $1+q$ & $q^{1/2} + 2q^{3/2}$ & $1+2q$ & $q^{1/2}$ & $1+2q$ & $q^{1/2}$ 
\end{tabular}
\end{center}

This finally completes the proof of the correspondence between the functions
$X_{2k+1} (a;b,c,d,e)$ and $y^{(ABC)}(k,k,0;l_{2},l_{3},l_{4})$ for $r=6$, as given in
section~\ref{sec:r6ident}.

\section{Conjectural expression for the local height probabilities: general case}
\label{sec:conjecture}

In the previous two sections, we proved the equality between the explicit
expressions (in terms of finitized CFT characters), and the functions $X_{2k+1}$,
in the cases $r=5$ and $r=6$. Now we present a general
conjecture for $X_{2k+1}$ for any $r \geq 5$. We start this section by motivating the form
of our conjecture.

The form of the fermionic formula for general $r$ closely resembles the fermionic formula for the height
probabilities of the original Andrews-Baxter-Forrester model, which were first conjectured in \cite{kkmcm93, kkmcm93a}, and
subsequently proven in several papers \cite{melzer94, berkovich94, s96,w96b}. 

The structure of these fermionic characters is that of the `Universal Chiral Partition Function' (UCPF), which can be
interpreted as a sum of states of several species of fermions \cite{bmc98}. This formalism has been developed in the nineties by
several groups, focussing on different aspects of the problem, such as the statistical mechanics models in which these
UPCFs appear \cite{dasmahapatra93}, representation theory \cite{lp85,kkmcm93, kkmcm93a},
or the connection with conformal field theory \cite{s97,gs00}.

For an introduction and more references, see the note \cite{bmc98}. Here, we would like to mention in particular the
paper \cite{bcr00}, in which the role of various types of particles which can appear in a UCPF is explained.
By making use of the connection with the central charge, we can find the `particle content' of the UCPF, which
in turn dictates its detailed form. 

In the UCPFs which describe the local height probabilities on the one hand, and the characters of the
CFT describing the critical behaviour on the other, two different types of particles appear, namely `real'
particles and so-called `pseudo' particles, which in effect have zero-energy and do not propagate. Their
presence does however effect the possible energies for the `real' particles. In addition, the presence of
`pseudo' particles has an effect on the central charge associated with the UCPFs, which is why we bring
up the notion of `pseudo' particles here.

We start by considering a set of particles, satisfying Haldane exclusion statistics \cite{h91},
with the statistics parameters encoded in the matrix $\bK$, where the elements $K_{ij}$ denotes the
mutual statistics between particles of type $i$ and $j$. The matrix ${\bf K}$ plays a central role in
the UCPF. The one-particle distribution functions
$\lambda_i$ for an ideal gas of fractional statistics particles were derived in
\cite{i94,do94,w94},
\begin{equation}
\label{iow}
\Bigl( \frac{\lambda_{i}-1}{\lambda_{i}}\Bigr) \prod_{j} \lambda_{j}^{K_{ij}} = z_i \ ,
\end{equation}
where $z_i = e^{-\beta \epsilon_i}$ is the fugacity of the species $i$. Introducing
$\lambda_{\rm tot} = \prod_{i} \lambda_{i} (z)$, where all the $\lambda_{i}$ are
evaluated at the same fugacity $z$, we can derive the central charge of the
associated CFT by making use of the relation between the central charge and the
specific heat. For the details, we refer to \cite{bcr00}. In short, the central charge associated
with a system of exclusion statistics particles characterized by $\bK$ (assuming that all particles
are real for now) is obtained by calculating the specific heat.
Specifically, one finds one has to solve the system of equations
\begin{equation}
\label{eq:ccalgebraic}
\xi_i = \prod_j (1-\xi_j)^{K_{ij}} \ .
\end{equation}
The solution $\{\xi_i\}$ is then used to calculate the central charge
\begin{equation}
c = \frac{6}{\pi^2} \sum_i L(\xi_i) \ ,
\end{equation}
where $L(z)$ is Rogers' dilogarithm
\begin{equation}
L(z) = -\frac{1}{2} \int_{0}^{z} dy \Bigl( \frac{\log y}{1-y} + \frac{\log(1-y)}{y} \Bigr) \ .
\end{equation}
For pseudo-particles, which keep track of the internal structure of the real particles,
we have $\epsilon_i = 0$, so $z_{i} =1$, which changes the $\lambda_{i}$ via
Eq.~\eqref{iow}. To calculate the central charge in the presence of pseudo-particles,
one first calculates the central charge associated with the full matrix $\bK$, giving
$c_{\rm full}$. In addition, one calculates the central charge associated with the part of the
matrix $\bK$ which only contains the pseudo-particles, which we denote by $c_{\rm pseudo}$.
The total central charge of the system is simply the difference of the two
\begin{equation}
c_{\rm CFT} = c_{\rm full} - c_{\rm pseudo} \ .
\end{equation}

The statistics matrix $\bK$ plays a central role in the UCPFs, which describe the partition functions
of statistical mechanics models, as well as the (chiral) characters conformal field theories. For the
various sectors in CFTs, both the matrix $\bK$ and the type of the particles (real or pseudo) is the same.
The forms of the functions $y$ for $r=5$ and $r=6$ take the form of a UCPF. 
The bilinear forms appearing in the exponent of $q$, can be written as $q^{\frac{1}{2} \bm \cdot \bK \bm}$
(see Eqns. \eqref{eq:ydefr5},\eqref{eq:ydefr6} for the cases $r=5$ and $r=6$).
We use the discussion above to obtain the correct form of the matrix $\bK$, and the number of
pseudo-particles necessary for the explicit expression for the functions $X_{2k+1}(a;b,c,d,e;q)$ for
arbitrary $r$, but consider the original ABF model first. 

In the case of the original ABF model, we know that the critical behaviour of the model in regime
III, i.e. when $p\rightarrow 0$ is given in terms of the minimal models $\mathcal{M}(r-1,r)$,
which have a coset description
$\frac{su(2)_1 \times su(2)_{r-3}}{su(2)_{r-2}}$ with $r\geq 4$ \cite{gko86}.
The matrix $\bK$ which appears in the bilinear form in
the UCPFs for these minimal models was found to take the form
$\bK = \frac{1}{2} {\bf A}_{r-3}$, where ${\bf A}_{r-3}$ is the Cartan matrix of $su(r-2)$, and is given via
its elements as $(A_{r-3})_{i,j} = 2 \delta_{i,j} - \delta_{|i-j|,1}$, with $i,j=1,\ldots,r-3$.
To calculate the central charge from this
matrix, we need the additional information that only the particle associated with the first row and column
is a real particle, all the other particles are pseudo-particles.

We quote the results of the calculation here, and refer to \cite{k95} for the details. The solution to the
equation \eqref{eq:ccalgebraic} for the matrix $\bK = \frac{1}{2} {\bf A}_{r-3}$
is given by
\begin{equation}
\xi_j = 1 - \frac{\sin \bigl( \frac{\pi}{r} \bigr)^2 }
{\sin\bigl(\frac{(r-2+j) \pi}{r} \bigr)^2} \ ,
\end{equation}
with $j=1,2,\ldots,r-3$. To obtain the central charge, dilogarithm identities were used, see \cite{k95}, which
resulted in the central charge associated with the matrix $\bK =\frac{1}{2} {\bf A}_{r-3}$
(assuming that all particles are real), namely
$c_{r} = \frac{6}{\pi^2} \sum_{j=1}^{r-3} L (\xi_j) = \frac{(r-3)(r-2)}{r}$.
Thus, the central charge of the minimal
models, for which only the first particle is real, is given by
$c_{\rm mm} = c_{r} - c_{r-1} = 1 - \frac{6}{(r-1)(r)}$, which is indeed the central charge of the minimal
model $\mathcal{M}(r-1,r)$.
We also note the following. The sum of the central charges associated with a matrix and its inverse
add up to the rank of the matrix (assuming that all particles are `real').
This means that the central charge associated with
$\bK^{-1} = 2 {\bf A}_{r-3}^{-1}$ is given by $r - 3 - \frac{(r-3)(r-2)}{r} = \frac{2(r-3)}{r}$, which is the
central charge associated with the $Z_{r-2}$-parafermion theory, which describes the critical
behaviour of the critical $p=0$ model for $u<0$, i.e. the critical behaviour associated with regime
II of the ABF model. We explain this connection in more detail below.

We now focus on the CFT description of the composite height model introduced in \cite{ka12}.
There, it was found that the critical behaviour in regime III was given
in terms of a diagonal coset model, also based on $su(2)_{r}$ affine Lie algebras,
in particular $\frac{su(2)_1 \times su(2)_1 \times su(2)_{r-4}}{su(2)_{r-2}}$. In \cite{ka12}, explicit
fermionic expressions for the local height probabilities, for finite system size, were found for
the special case where $r=5$. These local height probabilities took the form of
UCPFs, based on the matrix $\bK = \frac{1}{2} {\bf A}_2$, where both particles are real. The associated
central charge is indeed $c_5 = \frac{6}{5}$.

The central charge of the coset $\frac{su(2)_1 \times su(2)_1 \times su(2)_{r-4}}{su(2)_{r-2}}$
is given by $c = 1+1+\frac{3(r-4)}{r-2} - \frac{3(r-2)}{r} = 2 - \frac{12}{r(r-2)}$. By using the analogy
with the original ABF model, we can expect that the LHP of the composite model can be expressed
in terms of UCPFs, based on the matrix $\bK = \frac{1}{2} {\bf A}_{r-3}$, where two of the particles
are real, while the others are pseudo-particles. Indeed, the central charge associated with such a
UCPF is given by $c_{r} - c_{r-2} = 2 - \frac{12}{r(r-2)}$, which is the expected result. Below, we
will give the conjectured form of the LHP, written in terms of fermionic characters, based on the
bilinear form of $\bK = \frac{1}{2} {\bf A}_{r-3}$, where the first and last particles are real, while the
other particles are pseudo-particles.

\subsection{Conjectured fermionic form of the local height probability} \label{sec:conj_fermionic_form}

In this subsection, we will introduce the conjectured form of the functions $X_{r,2k+1} (a;b,c,d,e;q)$
which appear in the expression for the local height probability, for arbitrary value of $r$, thereby
generalizing the result given in \cite{ka12} for $r=5$, which we proved in the current paper, along with the case $r=6$. Note that we added the subscript $r$ to the notation $X_{r,2k+1} (a;b,c,d,e;q)$,
in order to be completely explicit. 

We first introduce a set of functions $\tilde{y}_{r}(k;l_2,l_3,l_4;q)$, which play a
central role in the description of the functions $X_{r,2k+1} (a;b,c,d,e;q)$.
In particular, we write
\begin{multline}
\label{eq:ytildedef}
\tilde{y}_{r}(k;l_2,l_3,l_4;q) =\sideset{}{'}\sum_{\substack{m_i \geq 0 \\ i=1,\ldots r-3}}
q^{\frac{1}{2} \bm \cdot {\bf K} \cdot \bm-\frac{1}{2}\delta_{1< l_4 < r-1} m_{r-1-l_4}}
\times \\
\qbinom{\frac{1}{2}(k+m_{2} + \delta_{l_3,1}+\delta_{l_4,r-2})} {m_1}
\biggl(
\prod_{i=2}^{r-4}
\qbinom{\frac{1}{2}(m_{i-1}+m_{i+1} + \delta_{l_3,i} + \delta_{l_4,r-1-i})}{m_i}
\biggr)
\qbinom{\frac{1}{2}(k+m_{r-4} + \delta_{l_2,1} + \delta_{l_3,r-3}+\delta_{l_4,2})} {m_{r-3}}
\ .
\end{multline}
Here, the labels $l_2$, $l_3$ and $l_4$ correspond to various factors in the coset
$\frac{su(2)_1 \times su(2)_1 \times su(2)_{r-4}}{su(2)_{r-2}}$, namely
the second factor $su(2)_1$, the factor
$su(2)_{r-4}$ and $su(2)_{r-2}$ in the coset for $l_2$, $l_3$ and $l_4$ respectively.
These labels take the values $l_2 = 1,2$,
$l_3 = 1,2,\ldots,r-3$ and $l_4 = 1,2,\ldots,r-1$. We have written the functions $\tilde{y}_{r}$ in such a
way that they are non-zero if $(l_2+l_3+l_4) \bmod 2 = 1$. This fixes the suppressed label $l_1$
corresponding to the first factor $su(2)_1$ in the coset to be $l_1 = (l_2 + l_3 + l_4 )\bmod 2$,
due to the constraint $l_1 + l_2 + l_3 = l_4 \bmod 2$. Finally, we introduced a `generalized
Kronecker-delta', $\delta_{\rm cond}$, which is $1$ if the condition $\textrm{`cond'}$ is met, and zero
otherwise.

The matrix $\bK = \frac{1}{2} {\bf A}_{r-3}$ has rank $r-3$ and is the Cartan matrix defined as above,
the prime on the sum denotes the constraint
that the summation variables are either even or odd, depending on the summation variable
as well as the other labels of the functions $\tilde{y}_{r}$. The parity of $m_i$ is given by
\begin{equation}
m_i \equiv
(k + l_3 + \delta_{i \geq r-l_4}\delta_{r-2+l_4+i\bmod2,0}
+ \delta_{i \leq l_3}\delta_{l_3+i\bmod2,1}) \bmod 2 \ .
\label{parities}
\end{equation}

With this, we have completely specified the functions $\tilde{y}_{r}$.
To make the connection with the functions $X_{r,2k+1}$, there is one additional step,
namely the introduction of the functions
$y_{r}(k;l_2,l_3,l_4;q)$, which differ from $\tilde{y}_{r}(k;l_2,l_3,l_4;q)$ only in the case that both
$l_2,l_3 > 1$. We define
\begin{align}
y_{r}(k;1,l_3,l_4;q) &= \tilde{y}_{r}(k;1,l_3,l_4;q) \\ \nonumber
y_{r}(k;2,l_3,l_4;q) &= \sum_{l=0}^{\min(l_3-1,2k+1)}
q^{(l (k + 1)/2 - (\bigl\lceil \frac{l^2-1}{2}\bigr\rceil)/4}
\tilde{y}_{r}(k-\bigl\lceil l/2\bigr\rceil;2-(l\bmod 2),l_3-l,l_4;q) \ ,
\label{eq:ydefgen}
\end{align}
where $\lceil x \rceil$ is the ceiling function, i.e. gives the smallest integer bigger or equal to $x$.
The structure of the sum in the case that $l_2=2$ is as follows. The argument $l'_4$ of the
functions $\tilde{y}_{r}(k';l'_2,l'_3,l'_4;q)$ is the
same in all the terms. The argument $l'_3$ decreases in steps of one starting from
$l'_3=l_3$. The argument $l'_2$ alternates between $l'_2=2$ and $l'_2=1$.
Finally, the argument $k'$ decreases by one every other term.
We note that the value $k'=-1$ can occur, in case that $2k \leq l_3-2$.

With the functions $y_{r}$ in place, we can now write down the explicit expressions for
the functions $X_{r,2k+1}$. We will do this for the
ground states for $u>0$ first. There are four types of ground state patterns, namely
$(b,b-1,b,b+1)$, $(b,b+1,b+2,b+1)$ and two others which are related by changing the
heights from $l$ to $r-l$, i.e. they are of the form $(b,b+1,b,b-1)$ and $(b,b-1,b-2,b-1)$.
Because the functions $X_{2k+1}$ remain unchanged when all the heights are reflected,
we will concentrate on the first two ground state patterns for $u>0$.
In particular, we have the following result
\begin{align}
X_{r,2k+1}(a;b,b-1,b,b+1) &= y_{r}(k;1,b-1,a;q)\\
X_{r,2k+1} (a;b,b+1,b+2,b+1) &= y_{r}(k;2,b,a;q) \ ,
\end{align}
where we note that the functions $y_{r}$ in the second line are a sum of terms $\tilde{y}_{r}$
in the case that $b\geq 2$.

The ground state patterns for $u<0$ are of the form $(b,b+1,b,b+1)$ or $(b,b-1,b,b-1)$.
By using the reflection symmetry \eqref{eq:refl} of the $X_{r,2k+1}$, we can
restrict ourselves to the first set of patterns.
As we explained in section \ref{sec:Xrecursion}, the functions $X_{r,2k+1}(a;b,b+1,b,b+1;q)$
are related to the ones for the ground state patterns for $u>0$, equations
\eqref{xreluneg1},\eqref{xreluneg2}. Namely, we have to distinguish two cases,
the functions $X_{r,2k+1} (a;b+1,b,b+1,b)$, where $b=1,2,\ldots,r-3$.
The remaining case $b=r-2$, i.e. $X_{r,2k+1} (a;r-1,r-2,r-1,r-2)$ will be
treated separately
\begin{align}
X_{r,2k+1} (a;b+1,b,b+1,b;q) &= q^{\frac{k+1}{2}} y_{r}(k;1,b,a;q) \\
X_{r,2k+1} (a;r-1,r-2,r-1,r-2;q) &= q^{k+1} y_{r}(k;2,1,r-a;q) \ .
\end{align}

Finally, we come to the functions $X_{r,2k+1}$ which are not related to ground state patterns,
namely $X_{r,2k+1} (a;b,b+1,b+2,b+3)$ and $X_{r,2k+1} (a;b+3,b+2,b+1,b)$, and concentrate on
the first one, where $b=1,2,\ldots,r-4$. As above, we use a relation to a ground state pattern
for $u>0$, namely \eqref{xrelngs}, to obtain
\begin{equation}
X_{r,2k+1} (a;b,b+1,b+2,b+3) = q^{\frac{k+1}{2}} y_{r}(k;2,b,a;q) \ .
\end{equation}

We numerically checked our conjecture for $r=5,\ldots, 13$, for all different ground
state patterns up to sizes of at least $k=10$ (in the case $r=13$). We believe that
it is possible to formalize the structure of the proof presented here for the cases
$r=5,6$, along the lines \cite{s96}, to prove the general case.

\section{Height probabilities in the thermodynamic limit}\label{sec:thermodynamic}

In this section, we take the thermodynamic limit of the obtained form for the
local height probabilities, in the case $u>0$ as well as for $u<0$. In the former
case, we can take the limit from the expressions \eqref{eq:ytildedef} directly, while
in the latter case, we first have to `invert' the expressions, in order to obtain the
result. For $u<0$, the result allows us to interpret the local height probabilities in
terms of CFT characters immediately. In the case $u>0$, this is more complicated,
and is the subject of the next section.

\subsection{Regime III $(u>0)$}

We start by considering the case $u>0$, and take the thermodynamic limit
$k\rightarrow\infty$, of the expressions for the height probabilities
$X_{r,2k+1} (a;b,c,d,e;q)$, which were expressed in terms of the functions
$y_{r} (k;l_2,l_3,l_4;q)$, given in eq. \eqref{eq:ydefgen}. In those cases where
both $l_2,l_3>1$, these expressions are a sum over various $\tilde{y}_{r}(k';l'_2,l'_3,l'_4;q)$, but
all terms, except for the first term $\tilde{y}_{r}(k;l_2,l_3,l_4;q)$, have an
overall factor $q^{l k/2}$, where $l$ is some positive integer. So in the thermodynamic
limit, we only have to consider the first term $\tilde{y}_{r}(k;l_2,l_3,l_4;q)$, which is
given in eq.~\eqref{eq:ytildedef}.

Taking the limit $\lim_{k\rightarrow\infty} \tilde{y}_{r}(k;l_2,l_3,l_4;q)$ is rather straightforward,
because the $k$ dependence only resides in the $q$-binomials, and we can 
make use of the result
$\lim_{n\rightarrow\infty} \qbinom{n}{m} = \frac{1}{(q)_m}$.
The only complication lies in the parity of the summation variables, which depends on
the parity of $k$, forcing us to take the limit $k\rightarrow\infty$ over either the even or
odd integers. In the end, we obtain the result
\begin{equation}
\label{eq:yinfinity}
\lim_{k\rightarrow\infty} y_{r}(k;l_2,l_3,l_4;q) = \\
\sideset{}{'}\sum_{\substack{m_i \geq 0 \\ i=1,\ldots r-3}}
\frac{q^{\frac{1}{2} \bm \cdot {\bf K} \cdot \bm-\frac{1}{2}\delta_{1< l_4 < r-1} m_{r-1-l_4}}}
{(q)_{m_1}(q)_{m_{r-3}}}
\prod_{i=2}^{r-4}
\qbinom{\frac{1}{2}(m_{i-1}+m_{i+1} + \delta_{l_3,i} + \delta_{l_4,r-1-i})}{m_i} \ ,
\end{equation}
where the limit is taken either over even or odd $k$, and the prime on the sum denotes
the constraints on the parities of the $m_i$, namely
\begin{equation}
m_i \equiv
(k + l_3 + \delta_{i \geq r-l_4}\delta_{r-2+l_4+i\bmod2,0}
+ \delta_{i \leq l_3}\delta_{l_3+i\bmod2,1}) \bmod 2 \ .
\label{parities2}
\end{equation}
The connection of these expressions with conformal field theory characters will be
given in the next section.

\subsection{Regime II $(u<0)$}
\label{sec:parafermions}

In order to obtain the thermodynamic limit of the local height probabilities for $u<0$, we
have to `invert' the characters. The reason is that in the regime $u>0$, the function
$\phi (\bl)$, see eq.~\eqref{eq:phidef}, has to be maximized to find the ground states.
Thus, to obtain the behavior at the
critical point, we need to do the following procedure. For any finite system size $k$, the
maximum value of $\phi (\bl)$ which can be obtained is $\frac{1}{2}(k+1)(k+2)$. This means
that the functions $X_{2k+1}$ we are interested in take the form
$q^{(k+1)(k+2)/2} X_{2k+1} (a;b,c,d,e;q^{-1})$, see also \cite{ka12} for more details.
The particular boundary conditions
relevant for the regime $u<0$ were described in section~\ref{subsec:phases}.
Equations~\eqref{xreluneg1} and \eqref{xreluneg2} express the
functions $X_{2k+1}$ for $u<0$ in terms of functions $X_{2k+1}$ relevant for
$u>0$. For the latter functions, we obtained the explicit fermionic expressions
$y_{r} (k;l_2,l_3,l_4;q)$, given in equations \eqref{eq:ytildedef} and \eqref{eq:ydefgen}, but
we note that for $u<0$, the relevant functions have either $l_2=1$ or $l_3=1$, so that we only need
to consider the functions $\tilde{y}_{r} (k;l_2,l_3,l_4;q)$ in eq.~\eqref{eq:ytildedef}.

For ease of notation, we write the functions $\tilde{y}_{r} (k;l_2,l_3,l_4;q)$ in the following
way
\begin{equation}
\tilde{y}_{r}(k;l_2,l_3,l_4;q) =\sideset{}{'}\sum_{\substack{m_i \geq 0 \\ i=1,\ldots r-3}}
q^{\frac{1}{2} \bm \cdot {\bf K} \cdot \bm- \frac{1}{2} {\bf B}\cdot \bm}
\prod_{i=1}^{r-3}
\qbinom{\bigl((1-{\bf K})\cdot \bm\bigr)_i + \frac{1}{2}({\bf k}+{\bf u})_i}
{m_i}
\ ,
\end{equation}
where the elements of the vectors ${\bf k}$, ${\bf u}$ and ${\bf B}$ are given by
$k_i = k(\delta_{i,1} + \delta_{i,r-3})$, 
$u_i = \delta_{i,r-3}\delta_{l_2,1} + \delta_{i,l_3} + \delta_{i,r-1-l_4}$
and 
$B_i = \delta_{i,r-1-l_4}$. The constraints on the sum are the same as those in
\eqref{parities2}.

We are interested in calculating $q^{(k+1)(k+2)/2} y_{r} (k;l_2,l_3,l_4;q^{-1})$, which
is done by making use of the identity
\begin{equation}
\qbinom{n+m}{m}_{q^{-1}} = q^{-nm} \qbinom{n+m}{m}_{q} \ , 
\end{equation}
and by changing the summation variables from the $m_i$ to
$n_i = \frac{1}{2} ({\bf k} + {\bf u})_{i} - ({\bf K}\cdot \bm)_i$.
Most of the steps
of this rewriting are straightforward. First, we note that we can write the $m_i$ in terms of
the $n_i$ as follows
$m_i = \bigl( {\bf K}^{-1} \cdot (\frac{1}{2} {\bf k} + \frac{1}{2} {\bf u} - {\bf n})\bigr)_i$, which
allows one to write the function $q^{(k+1)(k+2)/2} y_{r} (k;l_2,l_3,l_4;q^{-1})$
in terms of a sum over the $n_i$. One has to take care in taking into account the constraints
in the sum over the $m_i$ in the original expression. To this end, we consider the
sum $\sum_{i=1}^{r-3} i n_i$, by making use of the explicit form of the matrix ${\bf K}$,
whose elements read
${\bf K}_{i,j} =  \delta_{i,j} - \frac{1}{2} \delta_{|i-j|,1}$, where $i,j=1,\ldots,r-3$. In particular,
one finds $\sum_{i=1}^{r-3} i ({\bf K}\cdot \bm)_i  = \frac{r-2}{2} m_{r-3}$, giving rise to
$\sum_{i=1}^{r-3} i n_i = \frac{1}{2} \bigl((r-2)(k-m_{r-3}) + \sum_{i=1}^{r-3} i u_i \bigr)$.

The maximum value of $\sum_{i} i n_{i}$ is obtained for the minimal value $m_{r-3}$, which
is the parity of $m_{r-3}$, which we denote as $p_{r-3}$, i.e.
$p_{r-3} = (k + l_3 + \delta_{l_4\geq 3} \delta_{(l_4-1)\bmod2,0})\bmod 2$.
Thus, we find the following constraints
\begin{align}
\sum_{i=1}^{r-3} i n_i &\leq \frac{1}{2} \bigl((r-2)(k-p_{r-3}) + \sum_{i=1}^{r-3} i u_i \bigr) \nonumber \\
\sum_{i=1}^{r-3} i n_i &= \frac{1}{2} \bigl((r-2)(k-p_{r-3}) + \sum_{i=1}^{r-3} i u_i \bigr)\bmod r-2 \ ,
\label{eq:yinvconstraints}
\end{align}
where the second constraint follows from the fact that the parity of $m_{r-3}$ is fixed,
which means that $\sum_{i} i n_{i}$ can only go down in steps of $r-2$.

Expressing the remainder of the expression in terms of the $n_i$ is straightforward, and leads to
the following expression
\begin{align}
\label{eq:yinv1}
q^{\frac{1}{2} (k+1)(k+2)} y_{r} (k;l_2,l_3,l_4;q^{-1}) &= 
q^{\frac{(k+1)l_2}{2}}q^{\delta_{l_2,1}(\frac{r-1-2l_3}{4(r-2)})}
q^{-\frac{l_3(r-2-l_3)}{4(r-2)}}
q^{\frac{(l_4-1)(r-1-l_4)}{4(r-2)}} y^{\rm inv}_{r}(k;l_2,l_3,l_4;q) \\
y^{\rm inv}_{r}(k;l_2,l_3,l_4;q) &=
\sideset{}{'}\sum_{\substack{n_i \geq 0 \\ i=1,\ldots r-3}}
q^{\frac{1}{2}\bn \cdot {\bf K}^{-1} \cdot \bn - \frac{1}{2}{\bf B}\cdot {\bf K}^{-1} \cdot \bn}
\prod_{i=1}^{r-3}
\qbinom{k + \bigl((1- {\bf K}^{-1}) \cdot \bn \bigr)_i + \frac{1}{2} ({\bf K}^{-1}\cdot {\bf u})_i}
{n_i} \ ,
\label{eq:yinv2}
\end{align}
where the prime denotes the constraints \eqref{eq:yinvconstraints}, and we
note that the term 
$-\frac{1}{2}{\bf B}\cdot {\bf K}^{-1} \cdot \bn$ can be written as
$-\frac{1}{2}\delta_{2\leq l_4 \leq r-2} ({\bf K}^{-1} \cdot \bn)_{r-1-l_4}$.
Finally the terms $k$
in the binomials follow from $\frac{1}{2}({\bf K}^{-1} \cdot {\bf k})_i = k$, for all $i$.
The fact that all binomials contain the size dependence $k$ is of great importance. It means
that in the `inverted' character, all particles are real particles, no pseudo particles are
present anymore. Thus, in taking the thermodynamic limit $k\rightarrow\infty$, all
$q$-binomials are transformed in $\frac{1}{(q)_{n_i}}$. In this limit, the functions
$X_{2k+1}(a;b,c,d,e;q)$ relevant for the critical behavior in the regime $u<0$ 
correspond to the characters of the primary fields of the $Z_{r-2}$-parafermion
CFT.

To see this, we note that the first constraint in eq. \eqref{eq:yinvconstraints}
disappears in the limit $k\rightarrow\infty$. The second constraint requires a bit
more work. We first note that it is in fact independent of the size. Namely, $k$ has
the same parity as $p_{r-3}$ if $(l_3+\delta_{l_4\geq 3}\delta_{l_4-1 \bmod 2,0}) \bmod 2 = 0$, which
means we can write
\begin{equation*}
\sum_{i=1}^{r-3} i (n_i - \frac{1}{2}u_i) \bmod (r-2) = 
\begin{cases}
\frac{r-2}{2} & \text{if $(l_3 + \delta_{l_4 \geq 3} \delta_{l_4\bmod 2,1})\bmod 2 =1$} \\
0 & \text{otherwise}
\end{cases} \ .
\end{equation*}
To simplify, we use the result that
$\sum_{i} i u_i = l_3 + (r-1-l_4) \delta_{2\leq l_4 \leq r-2} + (r-3)\delta_{l_2,1}$. The range
on $\delta_{2\leq l_4 \leq r-2}$ can trivially be extended to $\delta_{2\leq l_4 \leq r-1}$.
The `exceptional case' $l_4=1$ in the expression for $\sum_{i} i n_{i}$ precisely allows
us to extent the range $\delta_{2\leq l_4 \leq r-1}$ to $\delta_{1\leq l_4 \leq r-1}$, i.e. the
full range of $l_4$, by making use of the relation $(l_2 + l_3 + l_4) \bmod 2 = 1$. Collecting all
the terms, we finally obtain the simple expression for the constraint
\begin{equation}
\sum_{i=1}^{r-3} i n_i = \frac{l_2 + l_3-l_4-1}{2} \bmod (r-2) \ .
\end{equation}
We can now make the connection between the functions $X_{r,2k+1}$ in the limit $k\rightarrow\infty$
and the characters of the primary fields of the $Z_{r-2}$ parafermion CFT. For details on this
theory, we refer to \cite{zf85}.

For $b=1,\ldots,r-3$, we obtain the following result
\begin{equation}
\begin{split}
&\lim_{k\rightarrow\infty} q^{\frac{1}{2} (k+1)(k+2)} X_{2k+1} (a;b+1,b,b+1,b;q^{-1}) =
\lim_{k\rightarrow\infty} q^{\frac{1}{2} (k+1)(k+2)} y_{r}(k;1,b,a;q^{-1}) = \\
& \lim_{k\rightarrow\infty}
q^{-\frac{(b-1)(r-b-1)}{4(r-2)}}
q^{\frac{(a-1)(r-a-1)}{4(r-2)}}
y_{r}^{\rm inv} (k;1,b,a;q) \propto \chq \Phi^{a-1}_{b-1}
\end{split}
\end{equation}
For $b=r-2$, we obtain the following expression
\begin{equation}
\begin{split}
&\lim_{k\rightarrow\infty} q^{\frac{1}{2} (k+1)(k+2)} X_{2k+1} (a;r-1,r-2,r-1,r-2;q^{-1}) =
\lim_{k\rightarrow\infty} q^{\frac{1}{2} (k+1)(k+2)} y_{r}(k;2,1,r-a;q^{-1}) = \\
& \lim_{k\rightarrow\infty}
q^{\frac{(a-1)(r-a-1)}{4(r-2)}}
y_{r}^{\rm inv} (k;2,1,r-a;q) \propto \chq \Phi^{a-1}_{r-3} \ .
\end{split}
\end{equation}
In the equations above, we denoted the fields of the $Z_{r-2}$ parafermion CFT
by $\Phi^{l}_{m}$, and the characters by $\chq \Phi^{l}_{m}$. These characters are
given by the following expressions (see, for instance \cite{lp85, kkmcm93, kkmcm93a, g95up, jacob02})
\begin{equation}
\chq \Phi^{l}_{m} =
\sideset{}{'}\sum_{\substack{n_i \geq 0 \\ i=1,\ldots r-3}}
\frac{q^{\frac{1}{2}\bn \cdot {\bf K}^{-1} \cdot \bn - \frac{1}{2}{\bf B}\cdot {\bf K}^{-1} \cdot \bn}}
{\prod_{i=1}^{r-3} (q)_{n_i}} \ ,
\label{eq:pfcharacter}
\end{equation}
where the prime denotes the constraint $\sum_{i=1}^{r-3} i n_i = \frac{l-m}{2} \bmod (r-2)$.
We find that in the limit $k\rightarrow\infty$, we obtain all the $Z_{r-2}$ characters
up to an overall factor of $q$. Namely, the labels $l,m$ of the fields $\Phi^{l}_{m}$ satisfy
the constraint $(l+m) \bmod 2 = 0$.
The label $m$ is taken to be modulo $2(r-2)$, because the fields
$\Phi^{l}_{m} \equiv \Phi^{l}_{m+2(r-2)}$ are identified. In addition, the fields
$\Phi^{l}_{m} \equiv \Phi^{r-2-l}_{m+r-2}$ are also identified \cite{zf85}. This implies that
we can restrict the labels $l$ and $m$ to the range $l=0,1,\ldots,r-2$ and
$m = 0, 1,\ldots,r-3$ for a total of $\frac{1}{2}(r-2)(r-1)$ fields. Thus identifications above
indeed cover all the fields in the $Z_{r-2}$ parafermion theory, establishing the
connection between the latter and the critical behavior of the composite height model for
$u<0$ (see also \cite{ka12}, noting that $\chq \Phi^{l}_{m} = \chq \Phi^{l}_{-m}$). 

Before we continue with the details of the connection between the functions $X_{2k+1}$
for $u>0$ and CFT characters, we would like to point out the following. In inverting the
characters above, we found that the pseudo-particles present in the fermionic description
of the functions $X_{r,2k+1}$ become real. The reason behind this was that upon inversion,
all the $q$-binomials acquired dependence on the size $k$, and therefore became factors
of $1/(q)_n$ in the limit $k\rightarrow\infty$. This is in fact the generic behavior. In the
original ABF model, the critical behavior for $u>0$ is given in terms of minimal models,
which have a fermionic description, with one real particle, while the other particles are
pseudo-particles. The matrix ${\bf K}$ was the same as the one used here. Upon inversion,
all the pseudo-particles become real. In fact, many (anyonic) chain models exhibit regions
whose critical behavior is described in terms of minimal models. Upon inverting the
sign of the hamiltonian, one often finds another critical region, whose criticality is described
in terms of the $Z_{r-2}$-parafermion theory. One can speculate that this behavior is governed
by integrable points, which exhibit the same behavior as observed for the ABF model, and
the composite height model considered here.

\section{Connection with a coset CFT: regime III $(u>0)$}
\label{sec:CFT}
In this section, we explain the connection between the expressions for the local
height probabilities in regime III $(u>0)$ and characters of a particular coset conformal field theory.
This connection for regime II $(u<0)$ was explained in the previous section.

To make the connection between the explicit UCPFs which we obtained for the functions $X_{r,2k+1}$, the local height
probabilities, and the CFT describing the critical behaviour of the model, we consider
bosonic forms of the characters associated with the coset conformal field theory
$\f{su(2)_{1}\times su(2)_{1}\times su(2)_{r-4}}{su(2)_{r-2}}$.

Before we start, we first collect a few properties of this coset theory. The fields $\Phi$ of this
theory carry, a priori, four labels, associated with the four $su(2)$ algebras. These four
labels, taking the values
$l_1,l_2 = 1, 2$, $l_3 = 1,\ldots, r-3$ and $l_4 = 1,\ldots,r-1$
satisfy the constraint $l_1+ l_2 + l_3 + l_4 = 0 \bmod 2$. Throughout, we make the choice
$l_1 = 1$, and frequently omit this label, and write the fields as $\Phi^{l_2,l_3}_{l_4}$.

The scaling dimensions $h_{l_2,l_3,l_4}$ of the fields were obtained in
\cite{ka12}, making use of the Coulomb gas results of \cite{ijs10}, in particular
\begin{equation}
h(l_{2},l_{3},l_{4})=\begin{cases}
\frac{(l_{3}r - l_{4}(r-2))^{2}-4}{8r(r-2)}+\frac{1}{2}-\frac{(l_{3}-l_{4}+2 l_{2})\bmod4}{4} & \text{for $l_{3}+l_{4}\bmod2=0$}\\
\frac{(l_{3}r - l_{4}(r-2))^{2}-4}{8r(r-2)}+\frac{1}{8} & \text{for $l_{3}+l_{4}\bmod2=1$}
\end{cases} \ .
\end{equation}
The scaling dimensions satisfy $h(3-l_{2},r-2-l_{3},r-l_{4})=h(l_{2},l_{3},l_{4})$,
reflecting the fact that the fields $\Phi_{r-l_{4}}^{3-l_{1},3-l_{2},r-2-l_{3}}$
and $\Phi_{l_{4}}^{l_{1},l_{2},l_{3}}$ are identified.

One way to view this coset theory is via a product of two unitary minimal models. The unitary minimal models have
a coset description, $\frac{su(2)_1\times su(2)_{r-3}}{su(2)_{r-2}}$ being the minimal model $\mathcal{M}(r-1,r)$, where $r=4$
corresponds to the Ising CFT. The coset $\f{su(2)_{1}\times su(2)_{1}\times su(2)_{r-4}}{su(2)_{r-2}}$ can be viewed
as the product of two minimal models, in particular
$\mathcal{M}(r-2,r-1) \ltimes \mathcal{M}(r-1,r) \cong
\frac{su(2)_1\times su(2)_{r-4}}{su(2)_{r-3}} \ltimes \frac{su(2)_1\times su(2)_{r-3}}{su(2)_{r-2}}$. 
The coset $\frac{su(2)_1 \times su(2)_1 \times su(2)_{r-4}}{su(2)_{r-2}}$ is not the direct product of the two consecutive minimal
models, but corresponds to a non-diagonal modular invariant of the product theory.
We note that in \cite{gs09}, the case $r=5$ was considered in the context of non-abelian quantum Hall states.

To construct this modular invariant, we will use the intuition that one can construct `new' CFTs from `old' ones by
condensing a boson (i.e., a particle corresponding to a field with integer scaling dimension) present in the theory,
as advocated in \cite{bs09}.
Under this condensation, the boson is identified with the vacuum, or identity field; in addition, those fields which are
related to each other by fusion with the boson (i.e., the condensate), are to be identified. Two other steps are necessary
to construct the new CFT. First, two fields which are each others dual (i.e., their fusion contains the identity) have
to be split, if both the identity and the boson which is condensed are present in the fusion product of the two fields considered.
This is because the boson is identified with the identity, and two fields can not be fused to the identity in more than one way.
Second, fields with have non-trivial monodromy with the boson, are confined. We used these principles to guide us in
constructing a non-diagonal modular invariant of the product theory of two consecutive minimal models.
More details about the condensation procedure can be found in \cite{bs09}.

To construct the relevant modular invariant, we first briefly recall some facts about minimal models \cite{bpz84}.
Minimal models  are labeled by two co-prime integers $r$ and $r'$, where we choose the ordering $r<r'$. The minimal model
$\mathcal{M}(r,r')$ is unitary if $r'=r+1$. The fields $\phi$ present in this theory can be labeled by two integers $(l_{3},l_{2})$, where
$1 \leq l_3 \leq r'-1$ and $1 \leq l_2 \leq r-1$. The scaling dimensions of the fields $\phi_{(l_3,l_2)}$ are given by
$h_{(l_3,l_2)} = \frac{(l_3 r - l_2 r')^2-(r-r')^2}{4 r r'}$. One finds that $h_{(l_3,l_2)} = h_{(r'-l_3,r-l_2)}$, and the labels
$(l_3,l_2)$ and $(r'-l_3,r-l_2)$ indeed correspond to the same field, $\phi_{(l_3,l_2)} = \phi_{(r'-l_3,r-l_2)}$.
Thus, the number of fields in the model $\mathcal{M}(r,r')$ is $\frac{1}{2}(r-1)(r'-1)$.

The chiral characters of the minimal models $\mathcal{M} (r,r')$ can be written as \cite{r85}
\begin{equation}
\chi_{(l_3,l_2)}^{(r,r')} (q) = \frac{q^{h-c/24}}{(q)_\infty} \sum_{n\in \mathbb{Z}} q^{n(n r r' + l_3 r - l_2 r')}-q^{(n r + l_2)(n r' + l_3)} \ ,
\end{equation}
where $c$ is the central charge of the corresponding theory, and $h$ the scaling dimension of the field.
In addition, $(q)_\infty = \prod_{n=1}^{\infty} (1-q^n)$.

The fields in the product theory $\mathcal{M}(r-2,r-1) \ltimes \mathcal{M}(r-1,r)$ are, in anticipation of
the results given below, labelled by
$(l_2,l_3;l_4,l'_2)$, and the scaling dimensions are given by the sum of the scaling dimensions of the
fields in the two factors. In particular, the field $(3,1;1,3)$ has scaling dimension $h_{(3,1;1,3)} = 2$, irrespective
of the value of $r$ (as long as $r\geq 5$, which is necessary for the coset to be defined). By applying the
condensation strategy
outlined above, one can find a set of fields which are not confined, and are inequivalent of one another.
This set of fields has the labels $(1,l_3;l_4,1)$ in the case that $l_3\bmod 2=1$ and $(2,l_3;l_4,2)$ when $l_{3}\bmod 2 = 0$.
We note that $l_3$ and $l_4$ take the values $l_3=1,2,\ldots,r-3$ and $l_4=1,2,\ldots,r-1$, which means that the product
theory contains $(r-3)(r-1)$ different fields. The fields in the original product theory which are identified with the
field with the label $(1,l_3;l_4,1)$ are of the form $(l,l_3;l_4,l)$, with $l \bmod 2 = 1$. The fields identified with
$(2,l_3;l_4,2)$ also take the form $(l,l_3;l_4,l)$, but now with $l \bmod 2 = 0$.

We will denote the characters of the field in the product theory by
$\chi_{(l_2,l_3;l_4,l'_2)}(q) = \chi_{(l_2,l_3)}(q) \chi_{(l_4,l'_2)}(q)$. With these fields,
one can construct different modular invariants. As usual, there is the diagonal invariant,
corresponding to the direct product of the two minimal models
\begin{equation}
Z_{\rm diag} = \sum_{(l_2,l_3;l_4,l'_2)=(1,1;1,1)}^{(r-2,r-3;r-1,r-2)} | \chi_{(l_2,l_3;l_4,l'_2)} (q) |^2 \ ,
\end{equation}
where the sum is over all the fields in the product theory.

Apart from this diagonal invariant, one can construct a different modular invariant, which corresponds to the
coset $\f{su(2)_{1}\times su(2)_{1}\times su(2)_{r-4}}{su(2)_{r-2}}$. We denote this theory by $\mathcal{M}(r-2,r-1,r)$,
and the partition function, which was obtained from the condensation picture outlined above, takes the form
\begin{equation}
Z_{\mathcal{M}(r-2,r-1,r)} =
\sum_{\substack{l_3=1\\ l_3 \;{\rm odd}}}^{r-3} \sum_{l_4=1}^{r-1} | \chi_{(1,l_3;l_4,1)} (q) + \chi_{(3,l_3;l_4,3)} (q) + \cdots|^2 +
\sum_{\substack{l_3=2\\ l_3 \;{\rm even}}}^{r-3} \sum_{l_4=1}^{r-1} | \chi_{(2,l_3;l_4,2)} (q) + \chi_{(4,l_3;l_4,4)} (q) + \cdots|^2 \ .
\end{equation}
This form of the modular invariant gives expressions for the characters of the coset $\mathcal{M}(r-2,r-1,r)$, in
terms of the bosonic characters of the minimal models.
The characters corresponding to the fields in the coset model can now be written in terms of the bosonic characters.

The characters of the coset fields are denoted as $\chq \Phi^{l_2,l_3}_{l_4}$ and are given by
\begin{equation}
\chq \Phi^{l_2,l_3}_{l_4} = \sum_{\substack{l=1\\ l+l_3 = 0 \bmod 2}}^{r-2} \chi_{(l,l_3;l_4,l)} (q) \ ,
\end{equation}
where it is assumed that $(l_2 + l_3 + l_4) \bmod 2 = 1$.
We checked numerically that these characters indeed correspond to the branching functions of the coset
$\mathcal{M}(r-2,r-1,r)$.

We can now relate the thermodynamic limit of the functions $X_{r,2k+1}$ to the characters of the coset fields. As already
indicated in \cite{ka12}, one has to consider the cases $k$ odd and $k$ even separately. The reason is that the
parity of the summation variables in the functions $\tilde{y}_{r}(k;l_2,l_3,l_4;q)$ depends on the parity of $k$.
We obtain the following identification. For $k=2p+1$ odd, the limits are, for the ground state patterns of type
$G_2^{-}$ and $G_1^{+}$, 
\begin{align}
\lim_{p\rightarrow\infty} X_{r,4p+3} (a;b+1,b,b+1,b+2;q) &= 
\lim_{p\rightarrow\infty} \tilde{y}_r (2p+1;1,b,a;q) = \chq \Phi^{1,b}_{a} \\ 
\lim_{p\rightarrow\infty} X_{r,4p+3} (a;b,b+1,b+2,b+1;q) &= 
\lim_{p\rightarrow\infty} \tilde{y}_r (2p+1;2,b,a;q) = \chq \Phi^{2,b}_{a} \\ \nonumber
&= \lim_{p\rightarrow\infty} \tilde{y}_r (2p+1;2,r-2-b,r-a;q) = \chq \Phi^{2,r-2-b}_{r-a} \ .
\end{align}
For $k=2p$ even, the connection between the functions $X_{r,2k+1}$ for the $u>0$ ground state patterns
$G_2^{-}$ is slightly different, namely
\begin{align}
\lim_{p\rightarrow\infty} X_{r,4p+1} (a;b+1,b,b+1,b+2;q) &= 
\lim_{p\rightarrow\infty} \tilde{y}_r (2p;1,r-2-b,r-a;q) = \chq \Phi^{1,r-2-b}_{r-a} \\ 
\lim_{p\rightarrow\infty} X_{r,4p+1} (a;b,b+1,b+2,b+1;q) &= 
\lim_{p\rightarrow\infty} \tilde{y}_r (2p;2,b,a;q) = \chq \Phi^{2,b}_{a} \\ \nonumber
&= \lim_{p\rightarrow\infty} \tilde{y}_r (2p;2,r-2-b,r-a;q) = \chq \Phi^{2,r-2-b}_{r-a} \ .
\end{align}
In taking the limit $k\rightarrow\infty$, the structure of the resulting expressions for $\tilde{y}_r$ 
is given in Eq. \eqref{eq:yinfinity}. The first and last $q$-binomials are transformed into $1/(q)_{m_1}$
and $1/(q)_{m_{r-3}}$. With this, the equivalence between the two different identifications for
the ground state patterns $G_1^{+}$ follows from a simple re-parametrization of the sum over the $m_{i}$.

\section{Discussion}

While the anyonic chains were introduced as a simple setting to study interacting anyons appearing in various topological phases, it has turned out that they have rich phase diagrams interesting in their own right, much like for ordinary spin chains. The two integrable critical points identified in Ref. \cite{ka12} led to the composite height model we have been studying here. In addition to novel phase diagrams and critical behavior, to mention one thing, the one-dimensional anyonic chains are of direct relevance to the boundary behavior of a nucleated phase, arising from the interactions, and the 'parent' topological phase hosting the bare anyons \cite{gat09}. By now, transitions between different topological phases due to the interactions between the anyonic excitations in topological liquids has been studied in quite some detail, see for instance \cite{gat09,gs09,bsh09}.

In this paper, we have further investigated the properties of the composite height model of Ref. \cite{ka12} as follows. We obtained and studied the fermionic forms of the LHPs and identified their off-critical CFT structure at two different regimes. The same CFTs are related to the critical points of the height model and that of the original anyon chain. The proof was based on the recurrence properties of the LHPs and the UPCFs, very much like in the original case of the ABF model and the minimal models. Proofs of this type, based on polynomial recurrences, are straightforward and tractable but unfortunately much of the physics, especially the off-critical CFT structure, in the height model is obscured as a trade off.

In particular, we gave a closed UCPF form for the LHPs. Using these, we were able to analytically prove the correspondence for $r=5,6$ and gave a general conjecture based on the structure of the UCPFs for $r=5,6$, correct asymptotic central charges, as well as numerical checks. Although the form of the recursion we used here quickly becomes cumbersome as $r$ grows larger, we suspect that our proof can be further improved to a proof for general $r$, along the lines of \cite{s96}, using the UCPFs conjectured here. We note that the fermionic representations of UCPFs as finitized characters are usually not unique \cite{bmc98}, as there might be several integrable perturbations of the CFT and different ways to introduce the `finitization' in the size $k$. We have not attempted to analyze to what integrable perturbations our UPCFs correspond to. Moreover, the recursions we were forced to use are more general than those in the lattice model, and the physical quantities related to the height probabilities were only obtained as special cases and, in the regime III, as linear combinations from the functions $y_{r=5,6}$. In addition, we obtained various relations between the general functions, which were necessary to show the equivalence with the local height probabilities. We note that in the case of the original ABF models, one can show more directly that the
recursions for the local height probabilities and the UCPFs are identical.

To the best of our knowledge, the fermionic forms for the characters of the coset $\f{su(2)_{1}\times su(2)_{1}\times su(2)_{r-4}}{su(2)_{r-2}}$
are new and characterized by the fact that they have two real fermions. In addition, the forms based on the characters of minimal models
also appears to be new. From these characters, we obtained characters of the $Z_{r-2}$ parafermions by sending $q\to q^{-1}$ and taking
the limit of large systems size in the obtained UPCFs. In the context of the anyonic chains, this procedure corresponds to changing the overall
sign of the hamiltonian. These infinite system size $Z_{r-2}$ parafermion characters are of course not new, but the finitized versions we obtained
here do differ from the finitized $Z_{r-2}$ parafermion characters which correspond to regime II in the original ABF model. The latter can be obtained
from the finitized characters corresponding to regime III of the ABF model by the inversion procedure, see for instance \cite{s96}. It is interesting that
the finitized characters of the diagonal coset models generically lead to different finitized characters of the $Z_{r-2}$ parafermions!

We briefly remark on the modular properties of the fermionic character formulae for the coset theories presented here. As our analysis in section \ref{sec:conjecture} shows, a crucial role is played by the pseudo-particles \cite{bcr00}. In the limit $k\to \infty$ for $u>0$, our fermionic formulae are generalized $r-3$-dimensional $q$-hypergeometric series containing $r-5$ $q$-binomial factors with finite arguments from the pseudo-particle sector. In the case when there are no pseudo-particles, Nahm \cite{nahm} has provided a conjecture regarding the modular properties of multi-dimensional $q$-hypergeometric series arising from UCPFs determined by a bilinear form $\bK$ and a `shift' ${\bf B}$. In the case $r=5$, there are no pseudo-particles present, and our formulae give back the well-known results and (asymptotic) central charges $c=\f{6}{5},\f{4}{5}$ based on $\bK=\f{1}{2}{\bf A}_{2}, 2{\bf A}_{2}^{-1}$ at rank 2 \cite{ka12, nahm}. In the general case for $u>0$, our fermionic character formulae \eqref{eq:yinfinity} have $q$-binomial contributions from the pseudo-particles, and the modular properties of the resulting $q$-series are even more complicated than in Nahm's conjecture and left for future study.

Lastly, we would like to mention some future directions for the study of the composite height model that were left out from this paper. Much like in the original ABF models, one would like to obtain bosonic forms for the characters of the coset theory, to pave way for corresponding Rogers-Ramanujan type identities and the modular properties of the coset theory. The UCPFs and their recursions that we have explicitly used in our proof for $r=5,6$ are more general than the LHPs and characters, that are obtained only at special values of the arguments and, for the coset theory, as linear combinations. This behavior is new compared to the ABF models and it would be interesting to find the natural representation theoretic setting, if any, of these $q$-polynomials or, conversely, to find a more direct functional form for the recursions in the LHPs. To study the physics and combinatorics of the height model more directly, a study of LHPs in terms of lattice paths of the composite height model \cite{feverati03, lamy-poirier11}, where \eqref{eq:phidef} acts as the Virasoro generator $L_{0}$, would be interesting. This would at the very least combinatorially relate generating functions in the path space of the height model to our fermionic UCPFs. Also, we have addressed only a half of the phase diagram of the composite height model, $0<p<1$, along the lines of the original paper \cite{ka12}, that corresponds to the quantum mechanical anyon chain. The negative $p$ regime, $-1< p< 0$, can also be studied via the CTM method but the quantum mechanical interpretation, if any, is unknown. The details of the phases of the composite height model for negative $p$ will be given in a subsequent publication \cite{kakashvili}.

\section*{Acknowledgements}
The authors thank the Institut Henri Poincar\'{e} for hospitality during the workshop
`Advanced conformal field theory and applications', at which the early stages of this
project were done.
J.N. also wishes to thank Nordita, where parts of his work were done, for a Visiting PhD Fellowship and great hospitality during his stay.
E.A. thanks P.~Kakashvili for discussions and collaboration on \cite{ka12,kakashvili}, and R.~Kedem for discussions and comments
on the manuscript.

\appendix

\section{Recursions for $r=6$}
\label{app:r6recursions}

Here we present the proof of the equivalence of the remaining 
recursions for $r=6$, one for each $a=l_{4}$ for the different patterns determined by the height $b$.
The other values for $a=l_4$ follow in almost the same way, since the patterns in the recursions and
the indices $l_{2},l_{3}$ are the same irrespective of $a=l_{4}$ and the identities for $y^{(ABC)}(k_{1},k_{2},k_{3};l_{2},l_{3},l_{4})$ are identical for fixed parity of $a=l_{4}$; thus the recursions differ only slightly in the specific parities appearing.

\subsection{Recursions in $G_1^{+}$}
The recursion for $X_{2k+1}(2;1232)$ is
\bea
X_{2k+1}(2;1232) = X_{2k-1}(2;1212) + X_{2k-1}(2;3212)
\eea
which is
\bean
X_{2k+1}(2;1232) = q^{k}y^{(BBB)}(k-1,k-1,0;212) + y^{(BBB)}(k-1,k-1,0;234) \\
+ q^{\f{k}{2}}y^{(AAB)}(k-2,k-2,0;124) +q^{\f{2k-1}{2}}y^{(ABA)}(k-2,k-2,0;214).
\eean
Now
\bean
y^{(ABC)}(k_{1},k_{2},k_{3};234) &=& y^{(CBA)}(k_{2},k_{1},k_{3};212)\\
y^{(ABC)}(k_{1},k_{2},k_{3};124) &=& y^{(CBA)}(k_{2},k_{1},k_{3}+1;212)\\
y^{(ABC)}(k_{1},k_{2},k_{3};214) &=& y^{(CBA)}(k_{2}-1,k_{1}+1,k_{3};212).
\eean
These give
\bean
X_{2k+1}(2;1232) = q^{k}y^{(BBB)}(k-1,k-1,0;212) + y^{(BBB)}(k-1,k-1,0;212) \\
+ q^{\f{k}{2}}y^{(BAA)}(k-2,k-2,1;212)+q^{\f{2k-1}{2}}y^{(ABA)}(k-3,k-1,0;212).
\eean
We use the relation in \eqref{eq:l4evenidentity},
\bean
y^{(ABC)}(k_{1},k_{2},k_{3};212) = y^{(C+1B+1A+1)}(k_{2}-1,k_{1}+1,k_{3};212), \quad \textrm{for } k_{3}\geq 0,
\eean
to get
\bean
X_{2k+1}(2;1232) = q^{k}y^{(BBB)}(k-1,k-1,0;212)+y^{(BBB)}(k-1,k-1,0;212) \\
+ q^{\f{k}{2}}y^{(BBA)}(k-3,k-1,1;212)+q^{\f{2k-1}{2}}y^{(ABA)}(k-3,k-1,0;212).
\eean
Now apply the recursion
\bean
y^{(BBA)}(k-3,k-1,1;212) = y^{(BBA)}(k-3,k-1,-1;212)+y^{(BAA)}(k-2,k,-1;212)
\eean
to get
\bean
X_{2k+1}(2;1232) = q^{\f{k}{2}}\left(y^{(BBA)}(k-3,k-1,-1;212)+q^{\f{k-1}{2}}y^{(ABA)}(k-3,k-1,0;212)\right)\\
+q^{k}y^{(BBB)}(k-1,k-1,0;212)+y^{(BBB)}(k-1,k-1,0;212)+q^{\f{k}{2}}y^{(BAA)}(k-2,k,-1;212).
\eean
This is equal to
\bean
X_{2k+1}(2;1232) = q^{\f{k}{2}}y^{(BBA)}(k-1,k-1,-1;212) + q^{k}y^{(BBB)}(k-1,k-1,0;212)\\
+y^{(BBB)}(k-1,k-1,0;212)+q^{\f{k}{2}}y^{(BAA)}(k-2,k,-1;212).
\eean
The RHS is equal to
\bean
q^{\f{k}{2}}\left(y^{(BBA)}(k-1,k+1,-1;212)+y^{(BAA)}(k-2,k,-1;212)\right)+y^{(BBB)}(k-1,k-1,0;212)\\
=q^{\f{k}{2}}y^{(BAA)}(k-2,k,1;212)+y^{(AAA)}(k-2,k,0;212),
\eean
where we use \eqref{eq:l4evenidentity} again on the second term. So we are left with
\bean
X_{2k+1}(2;1232) = q^{\f{k}{2}}y^{(BAA)}(k-2,k,1;212) + y^{(AAA)}(k-2,k,0;212) 
\eean
which is the original recursion for $X_{2k+1}(2;1232)=y^{(AAA)}(k,k,0;212)$. The recursion for $X_{2k+1}(4;1232)$ is similar and omitted.

Finally, the recursion for $X_{2k+1}(4;3454)$ is
\be
X_{2k+1}(4;3454) = q^{\f{k+1}{2}}X_{2k-1}(4;1234)+q^{\f{k+1}{2}}X_{2k-1}(4;3234)+X_{2k-1}(4;3434)+X_{2k-1}(4;5434).
\ee 
This is the same as
\bean
X_{2k+1}(4;3454) = q^{\f{k+1}{2}}q^{\f{k}{2}}y^{(BAB)}(k-1,k-1,0;214) + q^{\f{k+1}{2}}y^{(BBA)}(k-1,k-1,0;124)\\
+q^{\f{k}{2}}y^{(BAA)}(k-1,k-1,0;122)+y^{(BBB)}(k-1,k-1,0;212).
\eean
Next, we use the relations on the last the two terms, which are not part of $X_{2k+1}(4;3454)$,
\bean
y^{(ABC)}(k_{1},k_{2},k_{3};122) &=& y^{(CBA)}(k_{2}+1,k_{1}-1,k_{3}+1;234)\\
y^{(ABC)}(k_{1},k_{2},k_{3};212) &=& y^{(CBA)}(k_{2},k_{1},k_{3};234).
\eean
So we get
\bean
X_{2k+1}(4;3454) = q^{\f{k+1}{2}}q^{\f{k}{2}}y^{(BAB)}(k-1,k-1,0;214) + q^{\f{k+1}{2}}y^{(BBA)}(k-1,k-1,0;124) \\
+ q^{\f{k}{2}}y^{(AAB)}(k,k-2,1;234)+ y^{(BBB)}(k-1,k-1,0;234) \ .
\eean
Also, using the relation
\bean
y^{(ABC)}(k_{1},k_{2},k_{3};234) = y^{(C+1B+1A+1)}(k_{2}+1,k_{1}-1,k_{3};234), \quad \textrm{for } k_{3}\geq 0,
\eean
in \eqref{eq:l4evenidentity} on the last term, the recursion reduces to
\bean
X_{2k+1}(4;3454) = q^{\f{k+1}{2}}q^{\f{k}{2}}y^{(BAB)}(k-1,k-1,0;214) + q^{\f{k+1}{2}}y^{(BBA)}(k-1,k-1,0;124) \\
+ q^{\f{k}{2}}y^{(AAB)}(k,k-2,1;234)+ y^{(AAA)}(k,k-2,0;234)
\eean
which is simply the same as
\begin{equation*} 
 X_{2k+1}(4;3454) = q^{\f{2k+1}{2}}y^{(BAB)}(k-1,k-1,0;214) + q^{\f{k+1}{2}}y^{(BBA)}(k-1,k-1,0;124) + y^{(AAA)}(k,k,0;234), 
\end{equation*}
as desired. Again, the recursion for $X_{2k+1}(2;3454)$ is similar and left for the reader.

\subsection{Recursions in $G_2^{-}$}
The recursions for $X_{2k+1}(2;3234)$ is 
\bea
X_{2k+1}(2;3234) = q^{\f{k+1}{2}}X_{2k-1}(2;1232)+q^{\f{k+1}{2}}X_{2k-1}(2;3232) + X_{2k-1}(2;3432)+X_{2k-1}(2,5432).
\eea
Writing the RHS in terms of $y$'s gives
\bean
X_{2k+1}(2;3234) =  q^{\f{k+1}{2}}y^{(CCC)}(k-1,k-1,0;212)+q^{\f{k+1}{2}}q^{\f{k}{2}}y^{(CAA)}(k-1,k-1,0;122) \\
+y^{(CCA)}(k-1,k-1,0;124)+ q^{\f{k}{2}}y^{(CAC)}(k-1,k-1,0;214).
\eean
Using
\bean
y^{(ABC)}(k_{1},k_{2},k_{3};212) &=& y^{(ABC)}(k_{1}+1,k_{2}-1,k_{3}-1;122)\\
y^{(ABC)}(k_{1},k_{2},k_{3};124) &=& y^{(CBA)}(k_{2}+1,k_{1}-1,k_{3};122)\\
y^{(ABC)}(k_{1},k_{2},k_{3};214) &=& y^{(CBA)}(k_{2},k_{1},k_{3}-1;122),
\eean
we get
\bean
X_{2k+1}(2;3234) = q^{\f{k+1}{2}}y^{(CCC)}(k,k-2,-1;122)+y^{(ACC)}(k,k-2,0;122)+q^{\f{k}{2}}y^{(CAC)}(k-1,k-1,-1;122)\\
+q^{\f{k+1}{2}}q^{\f{k}{2}}y^{(CAA)}(k-1,k-1,0;122).
\eean
Next we use the recursion on the last two terms
\ben
q^{\f{k+1}{2}}y^{(CAA)}(k-1,k-1,0;122)+y^{(CAC)}(k-1,k-1,-1;122) = y^{(CAC)}(k-1,k+1,-1;122),
\een
so
\begin{gather*}
X_{2k+1}(2;3234) = q^{\f{k}{2}}\left(q^{\f{1}{2}}y^{(CCC)}(k,k-2,-1;122)+y^{(CAC)}(k-1,k+1,-1;122)\right) + y^{(ACC)}(k,k-2,0;122).
\end{gather*}
Focusing on the first term, we have by \eqref{eq:l4evenidentity} and the recursion
\begin{gather*}
q^{\f{1}{2}}y^{(CCC)}(k,k-2,-1;122) = q^{1/2}y^{(AAA)}(k-1,k-1,-1;122) =\\
q^{1/2}y^{(AAA)}(k+1,k-1,-1;122)-q^{\f{k+1}{2}}y^{(CAA)}(k-1,k-1,0;122)
\end{gather*}
and
\ben
y^{(CAC)}(k-1,k+1,-1;122) = y^{(ACA)}(k,k-2,-1;122) + q^{\f{k+1}{2}}y^{(CCA)}(k,k-2,0;122).
\een
Using \eqref{eq:l4evenidentity} again, two terms cancel and give
\ben
q^{1/2}y^{(AAA)}(k+1,k-1,-1;122)+y^{(ACA)}(k,k-2,-1;122) = y^{(ACA)}(k,k-2,1;122).
\een
Finally we get
\ben
X_{2k+1}(2;3234) =  q^{\f{k}{2}}y^{ACA}(k,k-2,1;122) + y^{(ACC)}(k,k-2,0;122) \ .
\een
The recursion for $X_{2k+1}(4;3234)$ is similar and omitted.

Finally, the recursion for $X_{2k+1}(3;4345)$ is
\bea
X_{2k+1}(3;4345) = q^{\f{k+1}{2}}X_{2k-1}(3;2343) + q^{\f{k+1}{2}}X_{2k-1}(3;4343) + X_{2k-1}(3;4543) \ .
\eea
In terms of $y$'s the RHS is
\bean
X_{2k+1}(3;4345) = q^{\f{k+1}{2}}\left(y^{(BAB)}(k-1,k-1,0;223) + q^{\f{k}{2}}y^{(AAB)}(k-2,k-2,0;113)\right) \\
+ q^{\f{k+1}{2}}q^{\f{k}{2}}y^{(BAA)}(k-1,k-1,0;133) + y^{(BBA)}(k-1,k-1,0;113).
\eean
Now
\bean
y^{(ABC)}(k_{1},k_{2},k_{3};223) &=& y^{(ABC)}(k_{1},k_{2}-2,k_{3}+1;133)\\
y^{(ABC)}(k_{1},k_{2},k_{3};113) &=& y^{(ABC)}(k_{1}+1,k_{2}-1,k_{3};133) 
\eean
and we get
\bean
X_{2k+1}(3;4345)  = q^{\f{k+1}{2}}y^{(BAB)}(k-1,k-3,1;133) + q^{\f{k+1}{2}}q^{\f{k}{2}}y^{(AAB)}(k-1,k-3,0;133)\\
+q^{\f{k+1}{2}}q^{\f{k}{2}}y^{(BAA)}(k-1,k-1,0;133)+y^{(BBA)}(k,k-2,0;133) \ .
\eean
Now, using the identity $y^{(ABC)}(k_{1},k_{2},k_{3};133) = y^{(CBA)}(k_{2}+2,k_{1}-2,k_{3};133)$ in \eqref{eq:r6resum} on the last term, gives
\begin{gather}
X_{2k+1}(3;4345)= y^{(ABB)}(k,k-2,0;133)  \nonumber \\  
+q^{\f{k+1}{2}}\left(y^{(BAB)}(k-1,k-3,1;133) + q^{\f{k}{2}}y^{(AAB)}(k-1,k-3,0;133)
+q^{\f{k}{2}}y^{(BAA)}(k-1,k-1,0;1133) \right) \ . \label{eq:allterms}  
\end{gather}
The term in the brackets is, using the recursion in the first term and \eqref{eq:r6resum} in the second,
\begin{gather*}
y^{(BAB)}(k-1,k-3,1;133) + q^{\f{k}{2}}y^{(AAB)}(k-1,k-3,0;133)
+q^{\f{k}{2}}y^{(BAA)}(k-1,k-1,0;1133) =\\
y^{(BAB)}(k-1,k-3,-1;133) + y^{(BBB)}(k,k-2,-1;133) +q^{\f{k}{2}}y^{(BAA)}(k-1,k-3,0;133)\\
+q^{\f{k}{2}}y^{(BAA)}(k-1,k-1,0;133),
\end{gather*}
or further, using eqs. \eqref{eq:r6resum} and \eqref{eq:133identity} on the last term,
\bean
y^{(BAB)}(k-1,k-1,-1;133) + y^{(BBB)}(k,k-2,-1;133) + q^{\f{k}{2}}y^{(BAA)}(k-1,k-1,0;133) \\
= y^{(BAB)}(k-1,k-1,-1;133) + y^{(BBB)}(k,k-2,-1;133) + q^{\f{k}{2}}y^{(AAB)}(k-1,k-1,0;133) 
\eean
or
\bean
y^{(BAB)}(k+1,k-1,-1;133) + y^{(BBB)}(k,k-2,-1;133) = y^{(BBB)}(k,k-2,1;133) \\ 
=y^{(ABA)}(k,k-2,1;133) \ ,
\eean
which follows again by \eqref{eq:133identity}. Returning to \eqref{eq:allterms}, we get back $y^{(ABB)}(k,k,0;133) = X_{2k+1}(3;4345)$ as desired. The recursions for $X_{2k+1}(1;4345)$ and $X_{2k+1}(5;4345)$ are similar and omitted.


\end{document}